\documentclass[review]{elsarticle}

\usepackage{lineno,hyperref}
\modulolinenumbers[5] 

\usepackage{amsmath,graphicx}

\usepackage{hyperref}
\usepackage{subfigure}
\usepackage{slashbox}
\usepackage{marginnote}
\usepackage{amssymb,amstext,amsmath}
\usepackage{float}
\usepackage{amsfonts}
\usepackage{bbm}
\usepackage{bm}
\usepackage{verbatim}
\usepackage{multirow}
\usepackage{makecell}
\usepackage{caption}
\usepackage{amsthm}

\usepackage{amscd,amsmath,amssymb,amsfonts,latexsym,mathrsfs}
\usepackage{setspace} 
\usepackage{graphics,epsfig}
\usepackage{stmaryrd}

\usepackage[english]{babel}
\usepackage{amsmath}
\usepackage{amsfonts}
\usepackage{amssymb}
\usepackage{bm}
\usepackage{mathrsfs}
\usepackage{subfigure}
\usepackage{float}
\usepackage[usenames]{color}

\usepackage{flushend}
\usepackage{verbatim}
\usepackage{tabularx}
\usepackage{multirow} 

\makeatother

\usepackage{rotating}

\newcommand{\x}{{\bf x}}

\renewcommand{\vec}[1]{\mathbf{#1}}

\begin{document}

\begin{frontmatter}

\title{Adaptive Independent Sticky MCMC algorithms}

\author{L. Martino$^\star$, R. Casarin$^\Diamond$, F. Leisen$^\dagger$, D. Luengo$^{\ast}$}

%\authorrunning{Short form of author list} % if too long for running head

\address{$^\star$ University of Helsinki; luca.martino@helsinki.fi \\
$^\diamond$  University Ca' Foscari (Venice); r.casarin@unive.it\\
$^\dagger$   University of Kent; fabrizio.leisen@gmail.com\\
$^{\ast}$   Universidad Politecnica de Madrid;  david.luengo@upm.es \\
}

%\tnotetext[mytitlenote]{Fully documented templates are available in the elsarticle package on \href{http://www.ctan.org/tex-archive/macros/latex/contrib/elsarticle}{CTAN}.}

%% Group authors per affiliation:
%\author{Luca Martino,David Luengp}
%\address{Dep. of Signal Theory and Communic., Universidad Carlos III de Madrid, Legan\'es (Spain).}
%\author{Luca Martino}
%\address{Dep. of Mathematics and Statistics, University of Helsinki, Helsinki (Finland).}
%\author{David Luengo}
%\address{ Dep. of Signal Theory and Communic.,Universidad Polit\'ecnica de Madrid, Madrid (Spain).}
%\author{M\'onica F. Bugallo}
%\address{Dep. of Electrical and Computer Eng., Stony Brook University,  NY (USA).}

%% or include affiliations in footnotes:
%\author[mymainaddress,mysecondaryaddress]{Elsevier Inc}
%\ead[url]{www.elsevier.com}

%\author[mysecondaryaddress]{Global Customer Service\corref{mycorrespondingauthor}}
%\cortext[mycorrespondingauthor]{Corresponding author}
%\ead{support@elsevier.com}

%\address[mymainaddress]{1600 John F Kennedy Boulevard, Philadelphia}
%\address[mysecondaryaddress]{360 Park Avenue South, New York}

\begin{abstract}
In this work, we introduce a novel class of adaptive Monte Carlo methods, called adaptive independent sticky MCMC algorithms, for efficient sampling from a generic target probability density function (pdf). The new class of algorithms employs adaptive non-parametric proposal densities which become closer and closer to the target as the number of iterations increases. The proposal pdf is built using interpolation procedures based on a set of support points which is constructed iteratively based on previously drawn samples. 
The algorithm's efficiency is ensured by a test that controls the evolution of the set of support points.
This extra stage controls the computational cost and the convergence of the proposal density to the target. Each part of the novel family of algorithms is discussed and several examples are provided.
 Although the novel algorithms are presented for univariate target densities, we show that they can be easily extended to the multivariate context within a Gibbs-type sampler.
The ergodicity is ensured and discussed. Exhaustive numerical examples illustrate the efficiency of sticky schemes, both as a stand-alone methods to sample from complicated one-dimensional pdfs and within Gibbs in order to draw from multi-dimensional target distributions.

\end{abstract}


\end{frontmatter}

\linenumbers

\section{Introduction}

%{\color{red} ANOTHER advantage: estimation of the normalizing constant of the target when the adaptation is complete... fare un numerical experiment with this stuff... (2D with pieces)}

Markov chain Monte Carlo (MCMC) methods \citep{Liu04b,Robert04}  are very important tools for Bayesian inference and numerical approximation, which are widely employed in signal processing \citep{Fitzgerald01,Doucet05MCsigpro} and other related fields  \citep{Liu04b,RobRos09}. %because they can generate samples from any target distribution available up to a normalizing constant.
%
%The standard MCMC techniques require the specification of a proposal distribution and produce a Markov chain that converges to the target distribution.
%
A crucial issue in MCMC is the choice of a proposal probability density function (pdf), as this can strongly affect the mixing of the MCMC chain when the target pdf has a complex structure, e.g., multimodality and heavy tails.
Thus, in the last decade, a remarkable stream of literature focuses on adaptive proposal pdfs, which allow for self-tuning procedures of the MCMC algorithms, flexible movements within the sample space and improved acceptance rates \citep{andrieu2008tutorial,Haario01}.

Adaptive MCMC algorithms are used in many statistical applications and different schemes have been proposed in the literature  \citep{andrieu2008tutorial,Haario01,RobRos09,Liang10}. There are two main families of methods: the first strategy consists in adapting the parameters of a parametric proposal pdf according to the past values of the chain \citep{Haario01}. However, even if the parameters are perfectly adapted, a discrepancy between target and proposal pdf persists (except for the ideal case that the parametric families of proposal and target coincides). A second strategy attempts to adapt the entire shape of the proposal density using non-parametric procedures \citep{Gilks95, MartinoA2RMS}. Although the construction of the proposal and ensuring the ergodicity is usually more complicatedw, the resulting algorithms can be extremely efficient. 

In this work, we describe a general framework for designing suitable adaptive MCMC algorithms with non-parametric proposal densities. First, we describe the different blocks forming the novel class of algorithms and then provide several specific examples in all cases. The proposal density is non-parametric and the construction procedure relies upon alternative interpolation strategies. The user can control the $L_1$ distance between the proposal and the target pdf (i.e., the convergence of the proposal to the target) through the design of suitable statistical update test, which also controls the overall computational cost.

After describing the general features of the novel class, we introduce the adaptive independent sticky Metropolis (AISM) algorithm to draw efficiently from any (bounded) univariate target distribution.\footnote{The adjective ``sticky'' highlights the ability of the proposed schemes to generate a sequence of proposal densities that progressively ``stick'' to the target.}
Then, we also propose a more efficient scheme, called adaptive independent sticky Multiple Try Metropolis (AISMTM). The MTM technique \citep{Liu00} is an extension of the Metropolis-Hastings method, which fosters the exploration of the state space  \citep{CraiuLam07,MartinoJesse}.
Moreover, the new class of methods encompasses different well-known algorithms given in literature: the {\it Griddy Gibbs sampler} \citep{ritter1992griddyGibbs}, the {\it adaptive rejection Metropolis Sampling} (ARMS) \citep{Gilks95,Meyer08}, and the {\it independent doubly adaptive Metropolis Sampling} (IA$^2$RMS) \citep{MartinoA2RMS,A2RMSicassp}. 

The ergodicity of the adaptive sticky MCMC methods is ensured and discussed. The underlying theoretical support is based on the approach introduced in \citep{Holden09}. It is also important to remark that, AISM and AISMTM also provides automatically an estimation of the normalizing constant of the target (a.k.a. {\it marginal likelihood} or {\it Bayesian evidence})  (since, with a suitable choice of the update test, the proposal approaches the target pdf almost everywhere). This is usually a hard task using MCMC methods \citep{Liu04b,Liang10,Robert04}. %\citep{Chib01, Kroese11}.

%%If the adaptation is not stopped, the convergence of the sequence of proposal pdfs to the target is . 

%\citep{Cai08}

The structure of the paper is as follows.
Section \ref{Sect_GeneralScheme} introduces the generalities of  sticky MCMC methods and the AISM scheme.
Sections \ref{StickyProp} and \ref{AlternSect} present the general properties, jointly with specific examples, of the proposal constructions and the update control tests. Section \ref{AlgSpec} discusses some special techniques belonging to the class of sticky methods and Section  \ref{sec:aismtm} introduces AISMTM. Section \ref{RelWorks} completes the description of the related works and Section \ref{RangeAppSection} highlights the range of applicability of the proposed methodologies.
Numerical  simulations are provided in Section \ref{Sect_Simul}.
Section \ref{Sect_Concl} contains some conclusions.
%
%{\color{red} Finally, the Appendices contain the proofs of the theorems formulated in Sections \ref{StickyProp} and \ref{AlternSect}, a discussion on alternative constructions for the proposal and a detailed description of the structural limitation of ARMS.}

%drawing candidates from a simpler proposal, $\pi(x)$, and accepting or discarding them according to some appropriate rule.
%

\begin{comment}
{\color{magenta}
Par mi David, solo apuntes nada imp
\begin{enumerate}
\item {\color{red} provide the code....} poner code (hacer el control de la distancia automatico ...) poner y preparar c—digo 
%\item ARS, ARMS, and IA$2$RMS, Grid Gibbs? as special cases... 

\item decir? que el MTM puede verse como una mejora del aquel paper del resampling within Gibbs.
\item doubly intractable distribution, other application
\item hacer control de toda la biblio...ver si he puesto todo
\item calcular $Z_i$
\item calcular $Z_tot$ dentro de Gibbs
\item poner ESS en simu y \underline{codigo}.... y hablar de las posible aplicaciones a doubly intractable distribution
\item within Gibbs: discussion about $S_i$....y possible strategies...
\end{enumerate}
}
\end{comment}

\begin{figure*}[t]
\centering
\centerline{
%\subfigure[]{ \includegraphics[width=4cm]{ARMS_fig2.jpg}}
%\subfigure[]{ \includegraphics[width=3.95cm]{ARMS_fig7.pdf}}
\includegraphics[width=12cm]{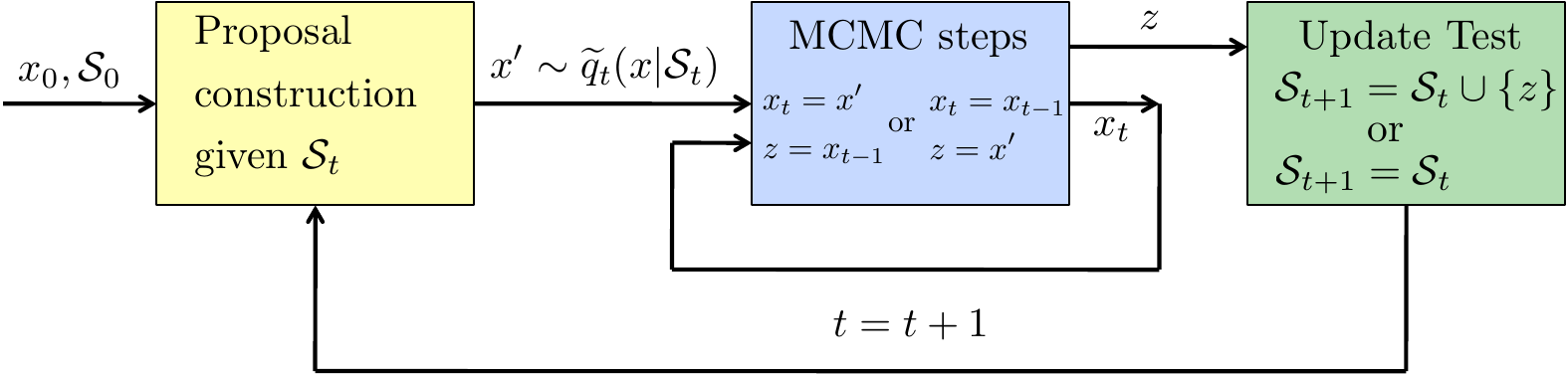}
}
\caption{{\footnotesize Graphical representation of a generic adaptive independent sticky MCMC algorithm.} }
\label{figCAZZOgeneral}
\end{figure*}
%%%%%%%%%%%%%%%%%%%%%%%%%%%%%%%%%%%%%
\section{Adaptive Sticky MCMC algorithms}\label{Sect_GeneralScheme}
%%%%%%%%%%%%%%%%%%%%%%%%%%%%%%%%%%%%%%

Let $\widetilde{\pi}(x) \propto \pi(x)>0$, with $x\in \mathcal{X}\subseteq \mathbb{R}$, be a bounded\footnote{For simplicity, we assume that $\pi(x)$ is bounded. However, the case of unbounded target pdfs can be tackled designing a suitable proposal construction, taking into account the vertical asymptotes of the target function.} target density known up to a normalizing constant, $c_\pi=\int_{\mathcal{X}} \pi(x) dx$, from which direct sampling is unfeasible. In order to draw from it, we employ an MCMC algorithm with an (independent) adaptive proposal density,
\begin{equation*}
	\widetilde{q}_t(x|\mathcal{S}_{t}) \propto q_t(x|\mathcal{S}_{t})>0, \quad x\in \mathcal{X},
\end{equation*}
where $t$ is the iteration index of the corresponding MCMC algorithm, and $\mathcal{S}_t=\{s_1,\ldots,s_{m_t}\}$ with $m_{t}>0$ is the set of support points used for building $\widetilde{q}_t$. An adaptive sticky MCMC method is conceptually formed by  three different stages:
\begin{enumerate}
\item {\it Construction of the non-parametric proposal:} given the nodes in $\mathcal{S}_t$,  
the function $q_t$ is built using a suitable non parametric procedure that provides a function which is closer and closer to the target as the number of points $m_t$ increases. 
\item {\it MCMC stage:}  steps of some MCMC method are applied in order to produce the next state $x_t$ if the chain, employing $\widetilde{q}_t(x|\mathcal{S}_{t})$ as proposal pdf.
\item {\it Update stage:} A statistical test is performed in order to decide whether to increase the number of points in $\mathcal{S}_t$ or not, defining a new set $\mathcal{S}_{t+1}$. This set $\mathcal{S}_{t+1}$ is used for constructing the proposal at the next iteration.
\end{enumerate}

Figure \ref{figCAZZOgeneral} provides a graphical sketch of a generic sticky MCMC method.
The update stage must be carefully designed. It has two important functionalities: controlling the computational cost and ensuring the ergodicity of the generated chain. See Appendix \ref{TeoApp} for some theoretical considerations.  Section \ref{StickyProp} describes the general properties that must be fulfilled by a suitable proposal construction, describing also several procedures for approximating the target $\widetilde{\pi}$ via interpolation. Section \ref{AlternSect} is devoted to the design of different suitable update rules.
As examples of MCMC structures for the second stage, in this work we consider a standard Metropolis-Hastings (MH) method, and a Multiple Try Metropolis (MTM) method.
In the following section we describe the simplest possible sticky method, obtained by using the MH algorithm, whereas in Section \ref{sec:aismtm} we consider a more sophisticated technique that employs the MTM.\footnote{Note that any other MCMC techniques could be used.}

%%%%%%%%%%%%%%%%%%%%%%%%%%%%%%%
\subsection{Adaptive independent sticky Metropolis (AISM)}
%%%%%%%%%%%%%%%%%%%%%%%%%%%%%%%
\label{sec:aism}
The simplest method belonging to the class of sticky MCMC is the {\it adaptive independent sticky Metropolis} (AISM) technique, outlined in Table \ref{alg1}. The proposal pdf $\widetilde{q}_t(x|\mathcal{S}_{t})$ changes along the iterations (see step \ref{ItemBuild1} of Table \ref{alg1}) following an adaptation scheme that relies upon a suitable interpolation given the set of support points $\mathcal{S}_t$ (see Section \ref{StickyProp}). %The, AISM t method is {\it adaptive} and ``{\it sticky}''  in the sense that the sequence of proposal functions $\{q_t\}_{t=1}^{\infty}$ progressively emulate the shape of the target, i.e., the proposal densities become closer and closer to the target as the number of iterations increases.
Step \ref{ItemUpdate1} of Table \ref{alg1} applies a statistical control for updating the set $\mathcal{S}_t$. The point $z$, rejected in the previous MH test,
can be added to $\mathcal{S}_t$  with probability 
\begin{equation}
P_a(z)=\eta_t(z,d_t(z)),
\end{equation}
where 
$$
 \eta_t(z,d): \mathcal{X}\times \mathbb{R}^+\rightarrow [0,1],
$$
is an increasing test function w.r.t. the variable $d$, such that $\eta_t(z,0)=0$, and
\begin{equation}
d=d_t(z)=\left|\pi(z)-q_t(z|\mathcal{S}_{t})\right|.
\end{equation}
is the point distance between $\pi$ and $q_t$ at $z$. The rationale behind this test is to use information from the target density in order to include in the support set only those points where the proposal pdf differs substantially from the target value at $z$. Note that since $z$ is always different from the current state (i.e., $z\neq x_t$ for all $t$), then the proposal pdf is {\it independent} from the current state according to Holden's definition \citep{Holden09} and thus the theoretical analysis is greatly simplified.

%

%Moreover, a suitable construction of the proposal leads to a probability of adding a new point that converges to zero.
%
%This implies that both the total number of points in the support set and the computational cost of building the proposals are kept bounded along the iterations, provided that $\eta(0)=0$.
%
%Different choices of $\eta$, which ensure a quick convergence of the proposal to the target, are presented in Section \ref{AlternSect}.

%we provide several interpolation methods based on a partition of the support of $\pi(x)$.

%{\it adaptive}

%{\it independent}

%{\it sticky}

\begin{table}[!t]
	\centering
	\caption{Adaptive Independent Sticky Metropolis (AISM)} 	\label{alg1}
	\vspace{-0.3cm}
    \begin{tabular}{|p{0.95\columnwidth}|}
    \hline
\small
For $t=0,\ldots,T-1$:
\begin{enumerate}
\item {\bf Construction of the proposal:} \label{ItemBuild1} Build a proposal function $q_t(x|\mathcal{S}_{t}) $  via a suitable interpolation procedure using the set of support points $\mathcal{S}_{t}$ (see Section \ref{StickyProp}). 
\item {\bf MH step:} \label{ItemMH1}
\begin{enumerate}
\item[2.1] Draw $x' \sim \widetilde{q}_t(x|\mathcal{S}_{t}) \propto q_t(x|\mathcal{S}_{t}) $.
\item[2.2] Set $x_{t}=x'$ and $z=x_{t-1}$ with probability 
\begin{equation*}
%\label{AlphaMH}
\alpha=\min\left[1,\frac{\pi(x')q_t(x_t|\mathcal{S}_{t})}{\pi(x_t)q_t(x'|\mathcal{S}_{t})}\right].
\end{equation*}
Otherwise, set $x_{t}=x_{t-1}$ and $z=x'$ with probability $1-\alpha$.
\end{enumerate}
\item \label{ItemUpdate1} {\bf Test to update $\mathcal{S}_t$:} Let $\eta_t(z,d): \mathcal{X}\times\mathbb{R}^+\rightarrow [0,1]$ be an increasing function w.r.t. the variable $d$, such that $\eta_t(z,0)=0$ and $\lim\limits_{d \to \infty} \eta_t(z,d)=1$. Then, set 
\begin{equation*}
\mathcal{S}_{t+1}=
\begin{cases}
 \mathcal{S}_t \cup \{z\}, & \mbox{with prob.} \ P_a(z)=\eta_t(z,d_t(z)),\\
 \mathcal{S}_t, & \mbox{with prob.} \ 1-P_a(z),
\end{cases}
\end{equation*}
where $d_t(z)=\left|\pi(z)-q_t(z|\mathcal{S}_{t})\right|$.
\end{enumerate} \\
%\end{algo}
%\hrule\vspace{5pt}
%\end{minipage}
%\end{center}
	\hline
	\end{tabular}
%	\label{tab:arms}
\end{table}
%\vspace*{-18pt}
%\end{figure*}

%
%Finally, it should be noted that Algorithm \ref{alg1} is a special case of the adaptive independent sticky MTM presented in the next section (see Algorithm \ref{alg2}), so the proof of its validity is contained in the proof given below for the adaptive independent sticky MTM technique. %and, therefore, it is not given here.

%{\color{red} convergence stuffs....explain the name ``independent'', ``sticky''}

%%%%%%%%%%%%%%%%%%%%%%%%%%%%%%
\section{Construction of the sticky proposals}
\label{StickyProp}
%%%%%%%%%%%%%%%%%%%%%%%%%%%%%%

%%%%%%%%%%%%%%%%%%
%\subsection{General properties}
%%%%%%%%%%%%%%%%%%
There are many alternatives available for the construction of a suitable {\it sticky proposal} (SP) pdf for sticky MCMC algorithms. Let us consider a set 
$$\mathcal{S}_t=\{s_1,\ldots,s_{m_t}\},$$
 of $m_{t}=|\mathcal{S}_t|$ support points, with $s_i\in \mathcal{X}$ for all $i=1,\ldots,m_t$. There are two properties that a sticky proposal construction must satisfy:
\begin{enumerate}
\item The proposal function is positive, $q_t(x|\mathcal{S}_{t})>0$, for all $x\in\mathcal{X}$ and $t\in \mathbb{N}$.
\item The $L_1$ distance between $\pi$ and $q_t$ vanishes to zero when the number of support points diverges, i.e., if $m_t \to \infty$ then
\begin{eqnarray*}
	D_1(\pi,q_t) = \|\pi- q_t \|_1 &=& \int_{\mathcal{X}}|\pi(z)-q_t(z|\mathcal{S}_{t})|dz \\
	&=&\int_{\mathcal{X}} d_t(z) dz \to 0.
\end{eqnarray*}
\item Samples can be drawn directly and easily from the resulting proposal $\widetilde{q}_t(x|\mathcal{S}_{t})\propto q_t(x|\mathcal{S}_{t})$ using some exact sampling procedure.
%The resulting proposal pdf $\widetilde{q}_t(x|\mathcal{S}_{t})\propto q_t(x|\mathcal{S}_{t})$ is easy to draw from.
\end{enumerate}

In this section, we provide some examples of constructions that approximate the target pdf by interpolating points that belong to the graph of the target function $\pi$.
The name ``sticky''  highlights the ability of the adaptation schemes to generate a sequence of proposal pdfs that converges to the target, thus allowing for a complete adaptation of the proposal pdf (i.e., a ``glutinous'' proposal that progressively ``sticks'' more and more to the target).

\begin{figure*}[t]
\centering
\centerline{
%\subfigure[]{ \includegraphics[width=4cm]{ARMS_fig2.jpg}}
%\subfigure[]{ \includegraphics[width=3.95cm]{ARMS_fig7.pdf}}
\subfigure[]{ \includegraphics[width=3.95cm]{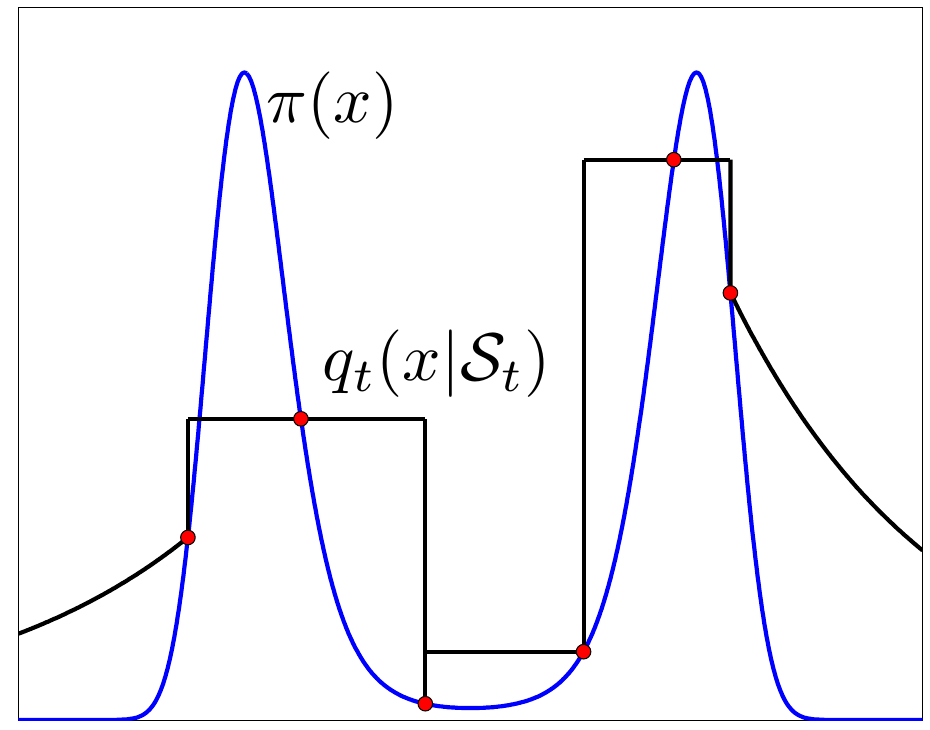}}
\subfigure[]{ \includegraphics[width=3.95cm]{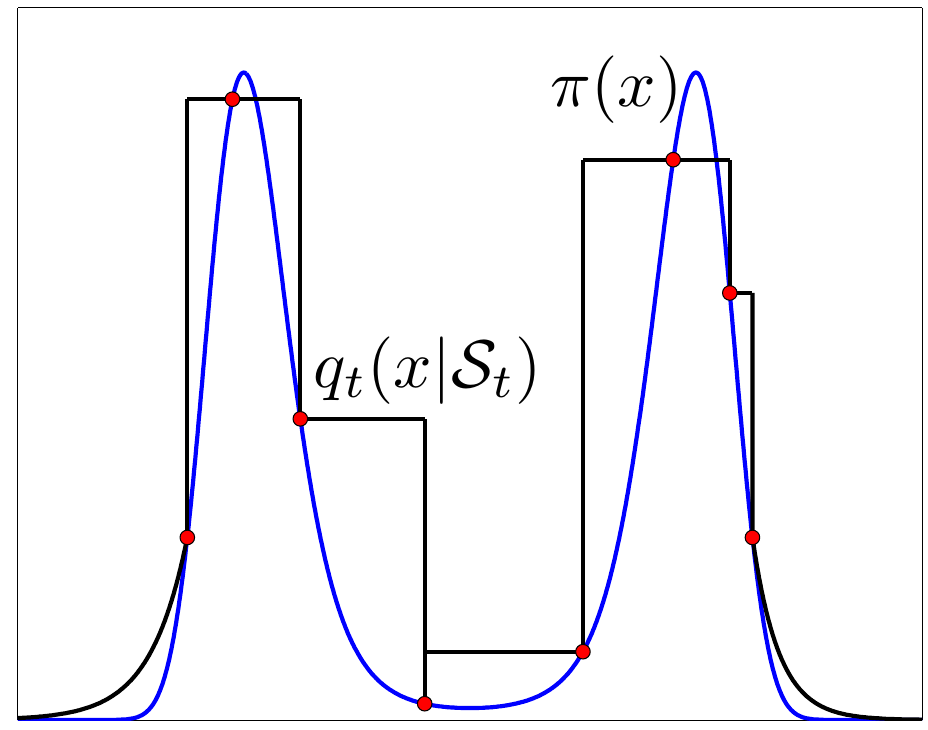}}
\subfigure[]{ \includegraphics[width=3.95cm]{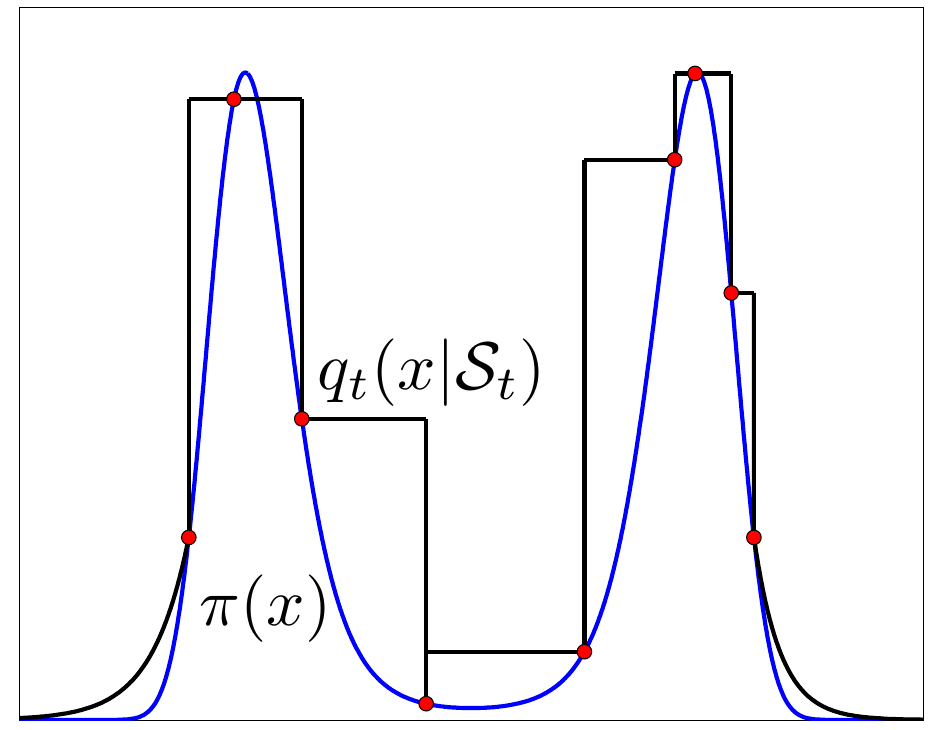}}
\subfigure[]{ \includegraphics[width=3.95cm]{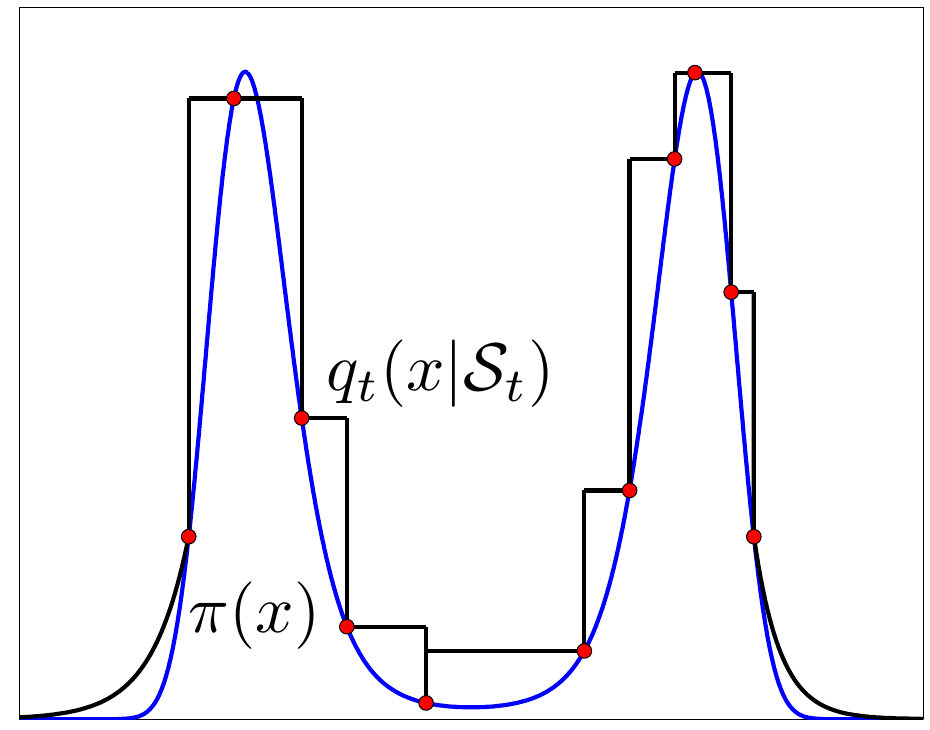}}
%%\subfigure[]{\includegraphics[width=4cm]{ARMS_fig10.jpg}}
}
\centerline{
%\subfigure[]{ \includegraphics[width=4cm]{ARMS_fig2.jpg}}
%\subfigure[]{ \includegraphics[width=3.95cm]{ARMS_fig7.pdf}}
\subfigure[]{ \includegraphics[width=3.95cm]{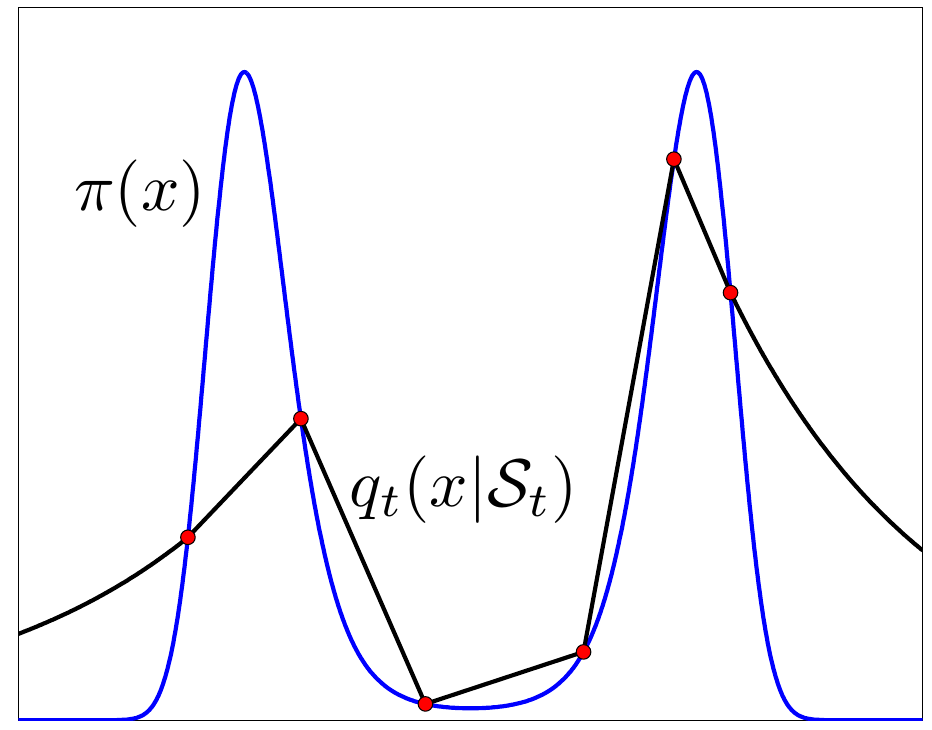}}
\subfigure[]{ \includegraphics[width=3.95cm]{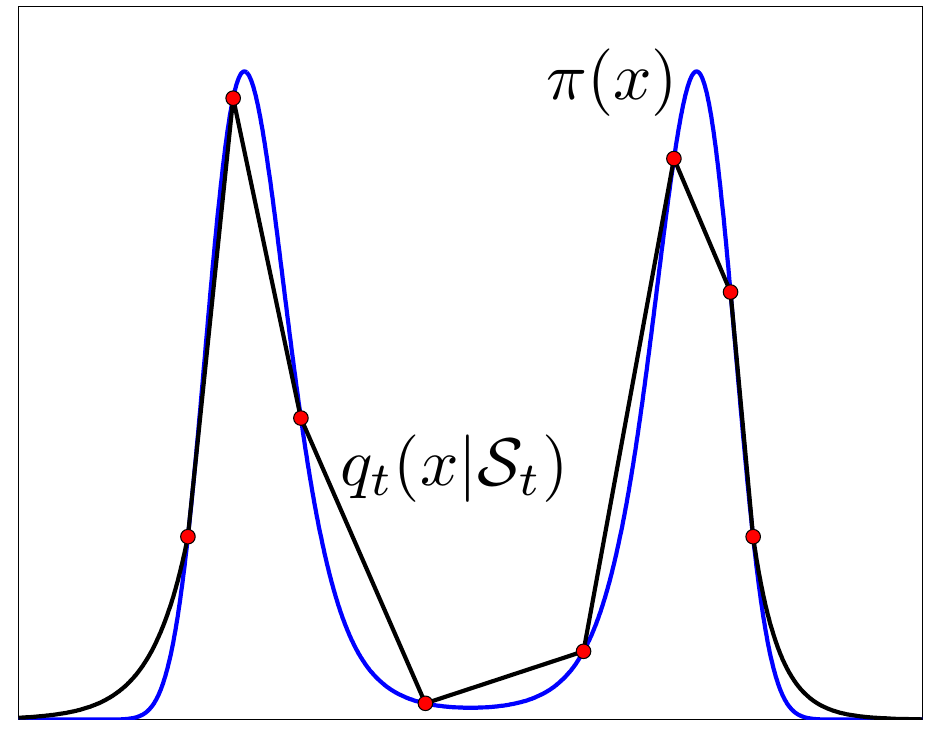}}
\subfigure[]{ \includegraphics[width=3.95cm]{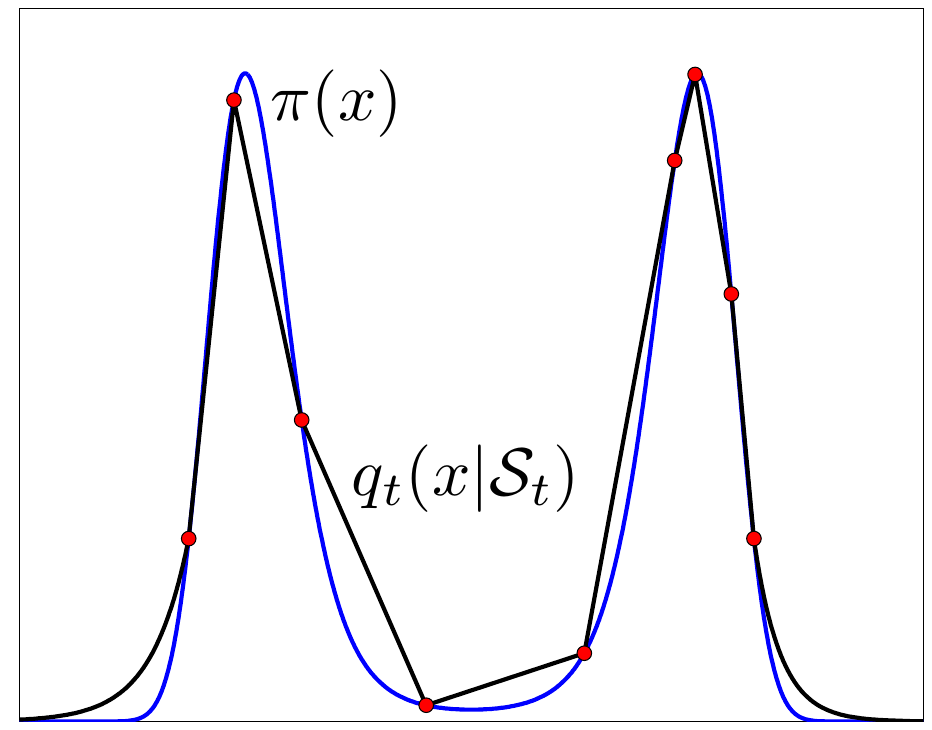}}
\subfigure[]{ \includegraphics[width=3.95cm]{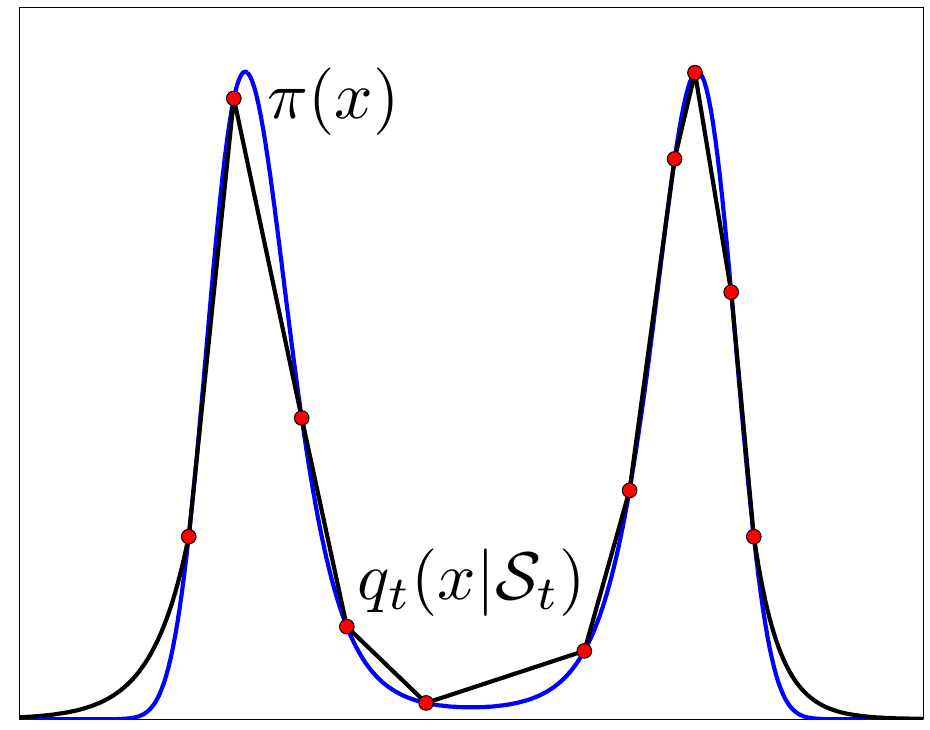}}
%%\subfigure[]{\includegraphics[width=4cm]{ARMS_fig10.jpg}}
}
\caption{Examples of the proposal construction $q_t$ considering a bimodal target $\pi$, using the procedures described in Eq.~(\ref{EqSimpler2}) for Figs. (a)-(b)-(c)-(d) and in Eq.~(\ref{EqSimpler3}) for Figs. (e)-(f)-(g)-(h) with $m_t=6,8,9,11$ support points, respectively.}
\label{figARMSProc}
\end{figure*}

%%%%%%%%%%%%%%%%%%%%%%%%
\subsection{Examples of constructions}
\label{sec:logDomain}
%%%%%%%%%%%%%%%%%%%%%%%%
Given $\mathcal{S}_t=\{s_1,\ldots,s_{m_t}\}$ at the $t$-th iteration, let us define a sequence of $m_t+1$ intervals: $\mathcal{I}_0=(-\infty,s_1]$, $\mathcal{I}_j=(s_j,s_{j+1}]$ for $j=1,\ldots,m_t-1$, and $\mathcal{I}_{m_t}=(s_{m_t},+\infty)$.
The simplest possible procedure uses  piecewise constant (uniform) pieces in $\mathcal{I}_i$, $1 \le i \le m_t-1$, with two exponential tails in the first and last intervals \citep{ritter1992griddyGibbs,FUSS,MartinoA2RMS}. %Indeed, the construction leads to the  proposal density formed by a collection of uniform pdfs with two exponential tails.
 More specifically, this can be mathematically defined as
\begin{gather}
\begin{split}
\label{EqSimpler2}
	W_{t}(x)= 
		\begin{cases}
			E_{0}(x), & x \in \mathcal{I}_0,\\
			\max\left\{\pi(s_{i}),\pi(s_{i+1}) \right\}, & x \in \mathcal{I}_i,\\
			E_{m_t}(x), & x \in \mathcal{I}_{m_t},
		\end{cases}
\end{split}
\end{gather} %}
where $1 \le i \le m_t-1$ and  $E_0(x)$, $E_{m_t}(x)$ represent two exponential pieces. These two exponential tails can be obtained simply constructing two straight lines in the log-domain as shown in \citep{Gilks92,Gilks95,MartinoA2RMS}. Other kinds of tails can be built, for instance using Pareto pieces (e.g., see \citep{MartinoA2RMS}). 
Alternatively,  we can use piecewise linear pieces \citep{Cai08}. %\footnote{Note that the structure of ATRAMS \citep{Cai08} is completely different to the AISM and ARMS-type techniques.}
 The basic idea is to build straight lines, $L_{i,i+1}(x)$, passing through the points $(s_{i},\pi(s_{i}))$ and $(s_{i+1},\pi(s_{i+1}))$ for $i=1,\ldots,m_t-1$, and two exponential pieces, $E_0(x)$ and $E_{m_t}(x)$, for the tails:
\begin{gather}
\begin{split}
\label{EqSimpler3}
	q_t(x|\mathcal{S}_t)=
		\begin{cases}
			E_{0}(x), &  \quad x \in \mathcal{I}_0,\\
			L_{i,i+1}(x), & \quad  x \in \mathcal{I}_i, \\
			E_{m_t}(x), & \quad  x \in \mathcal{I}_{m_t},\\
\end{cases}
\end{split}
\end{gather}
with  $i=1,\ldots,m_t-1$.
Unlike in \citep{Cai08}, here the tails $E_0(x)$ and $E_{m_t}(x)$ do not necessarily have to be equivalent in terms of the areas they enclose.
%
%
%
%Fig. \ref{figARMSProc}(d) depicts an example of the construction of $q_t(x|\mathcal{S}_t)$ using this last procedure.
%
Note that drawing samples from these trapezoidal pdfs inside $\mathcal{I}_i=(s_i,s_{i+1}]$ is straightforward \citep{Cai08,Hormann03}. Figure \ref{figARMSProc} shows examples of the construction of $q_t(x|\mathcal{S}_t)$ using Eq. \eqref{EqSimpler2} or \eqref{EqSimpler3} with different number of points, $m_t=6,8,9,11$.

A more sophisticated and computational expensive construction has been proposed for the ARMS method in \citep{Gilks95}. In this case, the proposal is formed by exponential pieces. Other alternative procedures can be found in the literature \citep{Gilks92,Meyer08,MartinoA2RMS,Cai08,FUSS}. A similar construction based on b-spline interpolation methods has been proposed in \citep{Krzykowski06,Shao13} for building a non-adaptive random walk proposal pdf for an MH algorithm. %A construction based on delta functions is proposed in \cite{}.  

%%%%%%%%%%%%%%%%%%%%%%%%%%%%%
\section{Update of the set of support points}
\label{AlternSect}
%%%%%%%%%%%%%%%%%%%%%%%%%%%%%

\begin{figure*}[t]
\centering
\centerline{
%\subfigure[]{ \includegraphics[width=4cm]{ARMS_fig2.jpg}}
%\subfigure[]{ \includegraphics[width=3.95cm]{ARMS_fig7.pdf}}
\includegraphics[width=10cm]{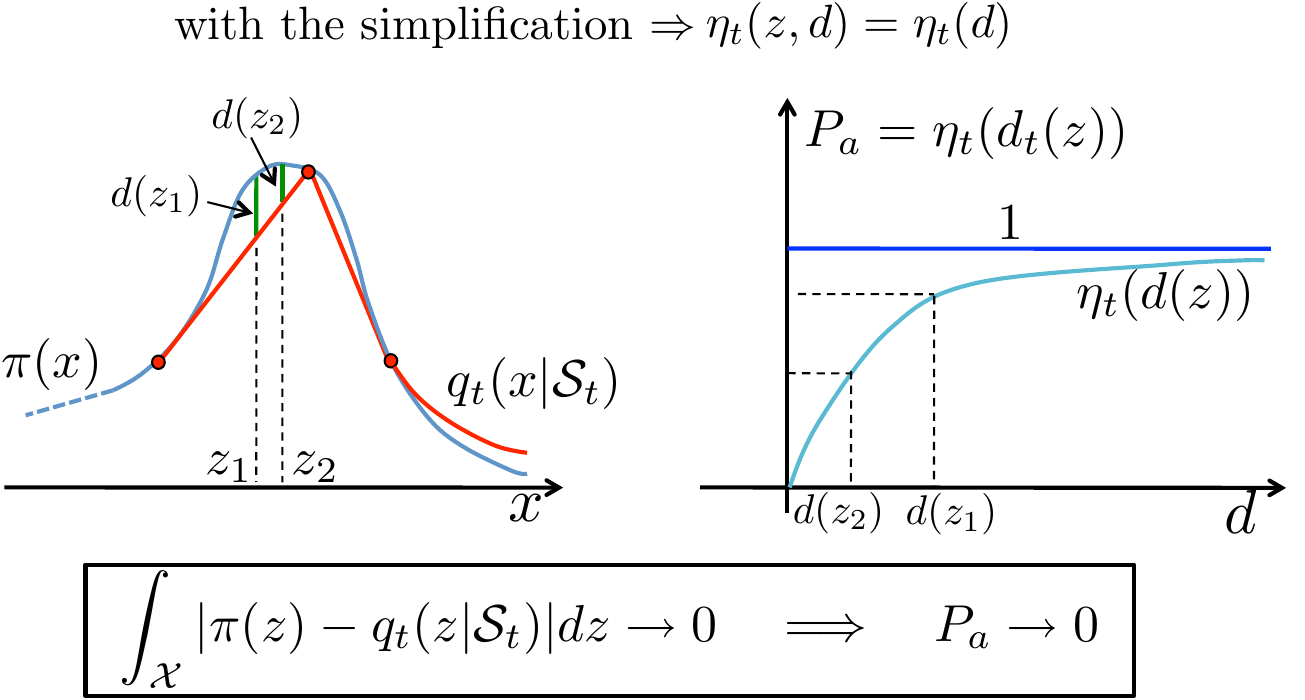}
}
\caption{{\footnotesize Graphical representation of the underlying idea behind the update control test. For simplicity, in this figure we have assumed $\eta_t(z,d)=\eta_t(d)$. As the proposal function $q_t$ becomes closer and closer to $\pi$, the probability of adding a new node to $\mathcal{S}_t$ decreases.} }
\label{figGenIDEA_RULE}
\end{figure*}

In AISM, a suitable choice of the function $\eta_t(z,d)$ is required. Any valid test function $\eta_t(z,d)$ must fulfill the following general properties: 
\begin{enumerate}
\item $\eta_t(z,d): \mathcal{X}\times \mathbb{R}^+\rightarrow [0,1]$,
\item $\frac{\partial \eta_t(z,d) }{\partial d}\geq 0$,  i.e., it is an increasing function w.r.t. the variable $d$, and
\item  $\eta_t(z,0)=0$.  
\item $\lim\limits_{d\rightarrow \infty}\eta_t(z,d)=1$.
\end{enumerate}
Figure \ref{figGenIDEA_RULE} depicts an example of function $\eta_t$ when $\eta_t(z,d)=\eta_t(d)$. Note that, for a given value of $z$, $\eta_t$ satisfies all the properties of a continuous distribution function (cdf) associated to a positive random variable. Therefore, any pdf for positive random variables can be used to define a valid test function $\eta_t$ through its corresponding cdf. In the following section we provide several examples of such test functions.

%%%%%%%%%%%%%%%%%%%%%%%%%%%
%\subsection{Key idea for sticky MCMC algorithms}
%%%%%%%%%%%%%%%%%%%%%%%%%%%
Recall that, given $d_t(z)=|\pi(z)-q_t(z|\mathcal{S}_t)|$, then $P_a(z)=\eta_t(z,d_t(z))$ is the probability of incorporating $z$ in $\mathcal{S}_t$.
Thus, %this update step can be seen as a way of measure of similarity between the proposal and target pdfs at a specific point $z$.
it is an important part of the algorithm, since it controls the trade-off between its performance and its computational cost  and jointly ensures the ergodicity of the chain.
Indeed, the use of a large number of support points improves the performance (as the proposal becomes closer to the target) at the expense of a higher storage and computational cost. Hence, a good adaptive strategy should only include new points only in regions where there is a large discrepancy between the proposal and the target functions. 

%Given the properties of $\eta_t$, another important observation is that the probability of adding a new point to $\mathcal{S}_t$ decreases if the distance $D_1$ diminishes (see App. \ref{}). Hence, the addition of a new point to $\mathcal{S}_t$, i.e., the increase of $m_t=|\mathcal{S}_t|$ has two joint effects:
%\begin{itemize}
%\item  the reduction the distance $D_1$ (given the properties of SPs).
%\item and, as a consequence, the reduction of  mean of the probability of $P_a(z)$.
%\end{itemize}
%Therefore, a suitable update rule manages the set $\mathcal{S}_t$ in order to build a ``good'' proposal with the smaller number $m_t$ possible, and keeping the ergodicity of the sampler. These concepts are summarized graphically in Figure \ref{figGenIDEA_RULE}.

%%%%%%%%%%%%%%%%%%%%%%%%%%%%%
\subsection{Examples of update rules}
%%%%%%%%%%%%%%%%%%%%%%%%%%%%%

Below, we provide different possible choices. 
First of all, we consider the simpler case of function of type
$\eta_t(z,d)=\eta(d)$.  
%In this case, the probability of adding a point $z$ is $P_a(z)=\eta(d_t(z))$ where $d_t(z)=|q(z|\mathcal{S}_t)-\pi(z)|$.
A first example, fulfilling the previous conditions, is
\begin{equation}
%\eta(d_t(z))&=\frac{1}{1+\exp\left\{-\gamma(d_t(z)-\varepsilon)\right\}}, \quad \quad d_t(z)&=|\pi(z)-q_t(z|\mathcal{S}_{t})|, \\
\eta_t(d)=1-e^{-\beta d}, 
\label{EqEsempio1bis}
\end{equation}
where $\beta>0$ is a constant parameter. Note that this is the cdc associated to an exponential random variable. A second possibility is
\begin{gather}
\eta_t(d)= \left\{
\begin{split}
&1, \quad \mbox{ if }\quad  d> \varepsilon,  \\
&0, \quad \mbox{ if }\quad  d\leq \varepsilon.  \\
\end{split}
\right. 
\label{EqEsempio1bis2}
\end{gather}
where $\varepsilon>0$ is pre-established parameter chosen by the user in advance. Since $P_a(z)=1$ if $ d_t(z)> \varepsilon$, this is a deterministic update rule where the number of support points is controlled through the threshold parameter, $\varepsilon$. Observe that  with $\varepsilon=\infty$ the update of $\mathcal{S}_t$ never happens whereas,  with $\varepsilon=0$ (this value is not allowed), the new node would be always incorporated to $\mathcal{S}_t$.\footnote{Regarding the selection of $\varepsilon$, note also that $d_t(z) \le \max\{\pi(z), q_t(z|\mathcal{S}_t)\} \le  M_{\pi}=\max\limits_{z\in \mathcal{X}}\{\pi(z)\}$, since $M_t=\max\limits_{z\in \mathcal{X}} q_t(z|\mathcal{S}_t)\le M_{\pi}$ for the described constructions. Then, $\epsilon$ can be chosen as a fraction of $M_\pi$, i.e., $\varepsilon =\kappa M_{\pi}$ with $0 < \kappa < 1$. Since $q_t$ becomes closer and closer to $\pi$, we have  $M_t\approx M_\pi$,  and this approximation improves with $t\rightarrow \infty$.} 
Moreover, with some  $0<\epsilon<1$, the adaptation could eventually stop and no support points would be added after some iterations.
 Eq. \eqref{EqEsempio1bis2} corresponds to the cdf associated to a Dirac's delta located at $\varepsilon$.
A third alternative is
\begin{equation}
\begin{aligned}
&\eta_t(z,d)=\frac{d}{\max\{\pi(z),q_t(z|\mathcal{S}_{t})\}}.\\
\end{aligned}
\label{EqFabriziobis}
\end{equation}
  Note that, since 
\begin{eqnarray}
d_t(z)&=&|\pi(z)-q_t(z|\mathcal{S}_t)|, \nonumber\\
&=&\max\{\pi(z),q_t(z|\mathcal{S}_{t})\}- \min\{\pi(z),q_t(z|\mathcal{S}_{t})\},\nonumber\\
&\le& \max\{\pi(z),q_t(z|\mathcal{S}_{t})\},
\end{eqnarray}
then $0\leq \eta_t(z,d)\leq 1$ for all $z$ and $d$. This rule appears in other related algorithms as we show below in Section \ref{AlgSpec}.
Furthermore, we can write $P_a(z)=\eta_t(z,d_t(z))$, with $\eta_t$ in Eq. \eqref{EqFabriziobis}, as
 \begin{equation}
\begin{aligned}
\eta_t(z,d_t(z))=1-\frac{\min\{\pi(z),q_t(z|\mathcal{S}_{t})\}}{\max\{\pi(z),q_t(z|\mathcal{S}_{t})\}}.
\end{aligned}
\label{EqFabriziobis_again}
\end{equation} 
This third rule corresponds to the cdf of a uniform random variable defined in the interval $[0,\max\{\pi(z),q_t(z|\mathcal{S}_{t})\}]$. Table \ref{UpdateRulesTable} summarizes the three previously described functions $\eta_t(z,d)$. %Clearly, several other choices of $\eta_t$ are possible.

 \begin{table}[hbt]
%\begin{sidewaystable}[p]
%\begin{table}[p]
\setlength{\tabcolsep}{2pt}
\def\marginwidth{1.5mm}
\begin{center}
%\label{resultsASM}
%\begin{tabular}{|l@{\hspace{\marginwidth}}|l@{\hspace{\marginwidth}}|}
\begin{tabular}{|l|l|}
%\begin{tabularx}{\textwidth}{|c|c|c|c|c|c|c|c|}
\hline
{\bf Rule 1} & $\eta_t(d)=1-e^{-\beta d}$   \\
\hline
\multirow{2}{*}{{\bf Rule 2} }& \multirow{2}{*}{
$\eta_t(d)=\Big\{$ 
 } $1,  \mbox{ if }  d> \varepsilon$, \\
  % $\eta_t(d)= 1, \mbox{ if }  d> \varepsilon$; \\
& \quad \quad \quad \quad \quad \hspace{0.05cm}$0,  \mbox{ if }  d\leq \varepsilon$   %   \mbox{ }
\\     
\hline                                                                          
{\bf Rule 3}  & $\eta_t(z,d)=\frac{d}{\max\{\pi(z),q_t(z|\mathcal{S}_{t})\}}$  \\                                                                                  
 \hline                                                                         
%\end{tabularx}
\end{tabular}
\end{center}
\vspace{-0.5cm}
\caption{ Examples of test function $\eta_t(z,d)$ for different update rules (recall that $d=d_t(z)=|q_t(z|\mathcal{S}_t)-\pi(z)|$). In the first and second cases, we have $\eta_t(z,d)=\eta_t(d)$.}\label{UpdateRulesTable}
%\end{sidewaystable} 
\end{table}

\section{Other examples of sticky MCMC methods}
%%%%%%%%%%%%%%%%%%%%%%%%%%%%%
\label{AlgSpec}

The novel class of adaptive independent MCMC methods encompasses several existing algorithms already available in the literature, as shown in Table \ref{SpecialTable}. We denote the proposal pdf employed in these methods as $p_t(x)$ and, for simplicity, we have removed the dependence on $\mathcal{S}_t$ in the function $q_t(x)$. The Griddy Gibbs Sampler \citep{ritter1992griddyGibbs} builds a proposal pdf as in Eq. \eqref{EqSimpler2}, which is never adapted later. ARMS \citep{Gilks95} and IA$^2$RMS \citep{MartinoA2RMS} use as proposal density
$$
p_t(x)\propto\min\{q_t(x), \pi(x)\},
$$
where $q_t(x)$ is built using different alternative methods \citep{Gilks95,Meyer08,FUSS,MartinoA2RMS}. Note that it is possible to draw easily from $p_t(x)\propto\min\{q_t(x), \pi(x)\}$ using the rejection sampling principle \citep{Tierney94,MartinoSigPro10}. ARMS adds new points to $\mathcal{S}_t$ using the update Rule 3, only when $q_t(z)\geq \pi(z)$, so that 
$$
P_a(z)=\eta_t(z,d_t(z))=1-\frac{\pi(z)}{q_t(z|\mathcal{S}_{t})}
$$
Otherwise, if  $q_t(z)< \pi(z)$, ARMS does not add new nodes (see the discussion in \citep{MartinoA2RMS} about the issues in ARMS mixing). Furthermore, the double update check used in IA$^2$RMS coincides exactly with Rule 3 when $p_t(x)\propto\min\{q_t(x), \pi(x)\}$ is employed as proposal pdf.  Finally, note that ARMS and IA$^2$RMS contain ARS in \citep{Gilks92} as special case when $q_t(x)\geq \pi(x)$, $\forall x\in \mathcal{X}$ and $\forall t\in \mathbb{N}$. Hence, ARS can be considered also a special case of the new class of algorithms. 

%note that the choice in Eq. \eqref{EqFabriziobis} resembles the probability of adding a new support point in the ARS method \citep{Gilks92}.
%
%Indeed, if $q_t(z|\mathcal{S}_{t}) \geq \pi(z)$ $\forall z\in\mathcal{X}$ and $\forall t$, then $\eta(d_t(z))=1-\frac{\pi(z)}{q_t(z|\mathcal{S}_{t})}$, which is exactly the probability of incorporating $z$ to the set of support points in ARS.

 \begin{table*}[hbt]
%\begin{sidewaystable}[p]
%\begin{table}[p]
\footnotesize
\setlength{\tabcolsep}{2pt}
\def\marginwidth{1.5mm}
\begin{center}
%\label{resultsASM}
%\begin{tabular}{|l@{\hspace{\marginwidth}}|l@{\hspace{\marginwidth}}|}
\begin{tabular}{|c|c|c|c|c|c|}
%\begin{tabularx}{\textwidth}{|c|c|c|c|c|c|c|c|}
\hline
{\bf Features} & {\bf Griddy Gibbs} &   {\bf ARMS} &  {\bf IA$^2$RMS} \\
\hline
\hline
Main Reference &\citep{ritter1992griddyGibbs} & \citep{Gilks95} & \citep{MartinoA2RMS}\\
\hline
Proposal pdf $p_t(x)$ & $p_t(x)=\widetilde{q}_t(x)$ &   $p_t(x)\propto\min\{q_t(x), \pi(x)\}$ & $p_t(x)\propto\min\{q_t(x), \pi(x)\}$  \\ 
\hline
Proposal Constr. & Eq. \eqref{EqSimpler2} & \citep{Gilks95},\citep{Meyer08}    & Eqs. \eqref{EqSimpler2}-\eqref{EqSimpler3},Ê\citep{MartinoA2RMS} \\    
 \hline    
\multirow{4}{*}{Update rule} &  never update, i.e., &If $q_t(z)\geq \pi(x)$ then Rule 3, & \multirow{4}{*}{Rule 3}   \\   
& Rule 2   &   If $q_t(z)< \pi(x)$ then  & \\       
&  with $\epsilon=\infty$ &   no update, i.e., & \\                                                                                                                                                      
&   &    Rule 2  with $\epsilon=\infty$  & \\                                                                                                                                                      
 \hline  
%\end{tabularx}
\end{tabular}
\end{center}
\vspace{-0.5cm}
\caption{Special cases of sticky MCMC algorithms. The ARS method in \citep{Gilks92} is a special case of ARMS and IA$^2$RMS, so that  ARS can be considered also belonging to the new class of techniques.
}\label{SpecialTable}
%\end{sidewaystable} 
\end{table*}

%%%%%%%%%%%%%%%%%%%%%%%%%%%%%%%%%
\section{Adaptive independent sticky MTM}
\label{sec:aismtm}
%%%%%%%%%%%%%%%%%%%%%%%%%%%%%%%%%%

In this section, we consider an alternative MCMC structure for the second stage described in Section \ref{sec:aism}: using a multiple-try Metropolis (MTM) approach \citep{Liu00,MartinoJesse}.  The resulting technique, Adaptive Independent Sticky MTM (AISMTM), is an extension of AISM that  considers multiple candidates as possible new state, at each iteration. This improves the ability of the chain to explore the state space \citep{MartinoJesse}. 
At iteration $t$, AISMTM builds the proposal density $q_t(x|\mathcal{S}_{t}) $ (step \ref{ItemBuild2} of Table \ref{alg2}) using the current set of support points $\mathcal{S}_t$.
Let $x_{t}=x$ be the current state of the chain and $x_{j}'$ ($j=1,\ldots,M$) a set of i.i.d. candidates simulated from $q_t(x|\mathcal{S}_{t})$ (see step \ref{ItemMTM2} of Table \ref{alg2}). 
 Note that, AISMTM uses an independent proposal (i.e., a non-random walk), just like AISM. As a consequence, the auxiliary points in step 2.3 of Table \ref{alg2} can be deterministically set \citep[pp. 119-120]{Liu04b},\citep{MartinoJesse}.

\begin{table}[!t]
	\centering
	\caption{Adaptive Independent Sticky Multiple Try Metropolis (AISMTM)} \label{alg2}
	\vspace{-0.3cm}
    \begin{tabular}{|p{0.95\columnwidth}|}
    \hline
\small
For $t=0,\ldots,T-1$:
\begin{enumerate}
\item {\bf Construction of the proposal:} \label{ItemBuild2} Build a proposal function $q_t(x|\mathcal{S}_{t})$ via a suitable interpolation procedure using the set of support points $\mathcal{S}_{t}$ (see Section \ref{StickyProp}).
\item {\bf MTM step:} \label{ItemMTM2}
\begin{enumerate}
\item[2.1] Draw $x_1',\dots,x_M'\sim \widetilde{q}_t(x|\mathcal{S}_{t}) \propto q_t(x|\mathcal{S}_{t}) $ and compute the weights $w_t(x'_i)=\frac{\pi(x'_i)}{q_{t}(x'_i|\mathcal{S}_{t})}$.
\item[2.2] Select $x'=x'_j$ among the $M$ tries with probability proportional to
$w_{t}(x'_i)$, for $i =1,\ldots, M$.
\item[2.3] Set the auxiliary point $x_i^*=x_i'$ and $z_i=x_i'$ for $i\neq j$. Moreover, set $x_j^*=x_{t-1}$.
\item[2.4] Set $x_{t}=x'$ and $z_j=x_{t-1}$ with probability 
\begin{equation*}
\label{AlphaMTM2}
\alpha=\min\left[1,\frac{w_t(x'_1)+\cdots+w_t(x'_M)}{w_t(x_1^*)+\cdots+w_t(x_M^*)}\right].
\end{equation*}
Otherwise, set $x_{t}=x_{t-1}$ and $z_j=x'$. 
\end{enumerate}
\item \label{ItemUpdateMTM2} {\bf Test to update $\mathcal{S}_{t}$:}  (see Section \ref{UpforAISMTM}) Select a point $z'$ within the set 
$\{z_1,\ldots, z_M\}$,
with probability proportional to some suitable weights $\varphi_t(z_i)$, for $i=1,\ldots,M$,  and set 
\begin{equation*}
\mathcal{S}_{t+1}=
\begin{cases}
 \mathcal{S}_t \cup \{z'\}, & \mbox{with prob.} \ P_a(z)=\eta_{t}(z',d_t(z')),\\
 \mathcal{S}_t, & \mbox{with prob.} \ 1-P_a(z),\\
\end{cases}
\end{equation*}
where $d_t(z)=|\pi(z)-q_t(z|\mathcal{S}_t)|$. For further information see  Section \ref{UpforAISMTM}.
\end{enumerate} \\
\hline
	\end{tabular}
%	\label{tab:arms}
\end{table}

In step \ref{ItemMTM2}, A sample $x'$ is selected among the set of candidates  $\{x_{1}',\ldots,x_{M}'\}$,
with probability proportional to the importance sampling weights,
\begin{equation*}
w_{t}(z)=\frac{\pi(z)}{q_{t}(z|\mathcal{S}_{t})}, \qquad \forall j \in \{1,\ldots,M\}.
\end{equation*}
The selected candidate is then accepted  or rejected according to the acceptance probability $\alpha$ given in step \ref{ItemMTM2}.
Finally, step \ref{ItemUpdateMTM2} updates the set $\mathcal{S}_t$,including a new point 
$$
z'\in\mathcal{Z}=\{z_1,\dots,z_M\},
$$
with probability $P_a(z')=\eta_t(z',d_t(z'))$. Note that $x_t\notin \mathcal{Z}$, and thus AISMTM is an independent MCMC algorithm according ot Holden's definition \citep{Holden09}.
For the sake of simplicity, we only consider the case where a single point can be added to $\mathcal{S}_t$ at each iteration. 
However, this update step can be easily extended to allow for more than one sample to be included into the set of support points.
Note also that AISMTM becomes AISM for $M=1$.

AISMTM provides a better choice of the new support points than AISM (see the numerical results). 
The price to pay for this increased efficiency is an higher computational cost per iteration.
However, since the proposal quickly approaches the target, it is possible to design strategies with a {\it decreasing} number of tries ($M_1 \ge M_2 \ge \cdots \ge M_t \ge \cdots \ge M_T$) in order to reduce the computational cost.

%%%%%%%%%%%%%%%%%%%%%%%%%%
\subsection{Update rules for AISMTM }
%%%%%%%%%%%%%%%%%%%%%%%%%%
\label{UpforAISMTM}
The update rules presented above require changes that take into account the multiple samples available, when used in AISMTM.
As an example, let us consider the update scheme in Eq. \eqref{EqFabriziobis}. %and let $z_{i}$ ($i=1,\ldots,M$) be a set of tries.
Considering for simplicity that only a single point can be incorporated to $\mathcal{S}_t$, the update step for $\mathcal{S}_t$ can be split in two parts: choose a ``bad'' point in $\mathcal{Z}\in \{z_1,\dots,z_M\}$ and then test whether it should be added or not.
Thus, first a $z'=z_i$ is selected among the samples in $\mathcal{Z}$  with probability proportional to
\begin{equation}
\label{PHIeq}
\begin{aligned}
\varphi_t(z_i)&=\max\left\{w_t(z_i),\frac{1}{w_t(z_i)}\right\} \\
&= \frac{\max\{\pi(z_i),q_t(z_i|\mathcal{S}_{t})\}}{\min\{\pi(z_i),q_t(z_i|\mathcal{S}_{t})\}},  \\
&= \frac{d_t(z_i)}{\min\{\pi(z_i),q_t(z_i|\mathcal{S}_{t})\}}+1, 
\end{aligned}
\end{equation}
for $i=1,\ldots,M$.\footnote{We have used the equality $d_t(z_i)=|\pi(z_i)-q_t(z_i|\mathcal{S}_t)|=\max\{\pi(z_i),q_t(z_i|\mathcal{S}_{t})\}-\min\{\pi(z_i),q_t(z_i|\mathcal{S}_{t})\}$.}
This step selects (with high probability) a sample where the proposal value is far from the target.
Then, the point $z'$ is included in $\mathcal{S}_t$ with probability 
\begin{eqnarray}
P_a(z')=\eta_{t}(z',d_t(z'))&=&1-\frac{1}{\varphi_t(z')}, \nonumber \\
&=&\frac{d_t(z')}{\max\{\pi(z'),q_t(z'|\mathcal{S}_{t})\}}, \nonumber
\end{eqnarray}
exactly as in Eq. \eqref{EqFabriziobis}.
Therefore, the probability of adding a point $z_i$ to $\mathcal{S}_{t}$ is
\begin{eqnarray*}
P_{\mathcal{Z}}(z_i)&=&\varphi_t(z_i)\eta_{t}(z_i,d_t(z_i)), \\
&=&\varphi_t(z_i)P_a(z_i)=\frac{\varphi_{t}(z_{i})-1}{\sum_{j=1}^{M}\varphi_{t}(z_{j})},
\end{eqnarray*}
that is a probability mass function defined over $M+1$ elements: $z_1$,$\ldots$, $z_M$ and the event $\{\mbox{\it no addition}\}$ that, for simplicity, we denote with the empty set symbol $\emptyset$.
Thus, the update rule in Step \ref{ItemUpdateMTM2} of Table \ref{alg2} can be rewritten as a unique step,  
\begin{equation}
\mathcal{S}_{t+1}=\left\{
%\begin{split}
\begin{array}{ll}
 \mathcal{S}_{t}\cup \{z_{1}\},  &\mbox{ with prob. } P_{\mathcal{Z}}(z_1)=\frac{\varphi_{t}(z_{1})-1}{\sum_{j=1}^{M}\varphi_{t}(z_{j})},\\
 &\vdots \\
 \mathcal{S}_{t}\cup \{z_{M}\},  &\mbox{ with prob. } P_{\mathcal{Z}}(z_M)=\frac{\varphi_{t}(z_{M})-1}{\sum_{j=1}^{M}\varphi_{t}(z_{j})},\\
 \mathcal{S}_t,  &\mbox{ with prob. }  \   P_{\mathcal{Z}}(\emptyset)=\frac{M}{\sum_{j=1}^{M}\varphi_{t}(z_{j})},\\
\end{array}
%\end{split}
\right.
\end{equation}
where we have used $1-\sum_{i=1}^M P_{\mathcal{Z}}(z_i)=\frac{M}{\sum_{j=1}^{M}\varphi_{t}(z_{j})}$.

%{\color{red} Other alternative rules for AISMTM can be designed following the ideas in Eqs. \eqref{EqEsempio1bis} and \eqref{EqEsempio1bis2}.
%%
%Moreover, several points can be included in $\mathcal{S}_t$ at the $t$-th iteration.
%%
%However, note that the proposal pdf should be rebuilt after each new addition in order to avoid an unnecessary increase of the number of points.
%%
%A possible black-box procedure would consist of adding points to the support set until a candidate is rejected.
%%
%This would allow us to control the number of support points added in each iteration automatically, since the probability of adding new support points would decrease as the proposal is updated.}

%%%%%%%%%%%%%%
\section{Related works}
%%%%%%%%%%%%%%
\label{RelWorks}
Other related methods, using non-parametric proposals, can be found in the literature. Samplers for drawing from univariate pdfs, using similar proposal constructions, has been proposed in \citep{Cai08,FUSS}, but the sequence of adaptive proposals does not converge to the target. Interpolation procedures for building the proposal pdf are also employed in \citep{Krzykowski06,Shao13}: however, in this case, the resulting proposal is a random walk-type (not independent) and the algorithm is not adaptive. Non-parametric proposal constructions have been also proposed for adaptive rejection sampling (ARS) \cite{Gilks92} and its extensions \citep{Gorur08rev,Hoermann95,MartinoSigPro10}. Other techniques have been  developed to be applied specifically for  ``Monte Carlo-within-in-Gibbs''  case where an importance sampling approximation of the univariate target pdf  is employed \citep{koch2007gibbs}.

%%%%%%%%%%%%%%%%%%%%%%%%%%%%%%%%%%%%%
\section{Range of applicability}
%%%%%%%%%%%%%%%%%%%%%%%%%%%%%%%%%%%%%%
\label{RangeAppSection}
The range of applicability  of the sticky MCMC methods is briefly discussed  below. On the one hand, sticky MCMC methods can be employed as stand-alone algorithms. Indeed, in many applications it is necessary to draw samples from complicated univariate target pdf (as example in signal processing, see \cite{Nakagami}). In this case, the sticky schemes provide virtually independent samples (i.e., with correlation close to zero) very efficiently. Moreover, at  the same time, they automatically give an approximation of the marginal likelihood.
 AISM and AIMTM can be also applied directly to draw from a multivariate distribution if a suitable construction procedure of the multivariate sticky proposal is designed (e.g, see \citep{Karawatzki06,Leydold98,Leydold98b,Hoermann95b}  and \cite[Chapter 11]{Hormann03}). However, devising and implementing such procedures in high dimensional state spaces are not easy tasks. Therefore, in this paper we focus on the use of the sticky schemes within other Monte Carlo techniques (such as Gibbs sampling or the hit and run algorithm) to draw from multivariate densities.
 %Section \ref{StickyProp} describe the general properties of a sticky proposal densities.   

%%%%%%%%%%%%%%%%%%%%%%%%%%%%%%%%%%%%%
\subsection{Sticky MCMC within other Monte Carlo method}
%%%%%%%%%%%%%%%%%%%%%%%%%%%%%%%%%%%%%%
%%%%%%%%%%%%%%%%%%%%%%%%%%%%%%%%%%%%%
%\subsubsection{Within a Gibbs-type algorithm}
%%%%%%%%%%%%%%%%%%%%%%%%%%%%%%%%%%%%%%
Bayesian inference often requires drawing samples from complicated multivariate posterior pdfs, $\widetilde{\pi}(\vec{x}|\vec{y})$ with 
$$
\vec{x}=[x_1,\ldots,x_L] \in \mathbb{R}^L, \quad L>1.
$$
For instance, this happens in blind equalization and source separation, or spectral analysis  \citep{Fitzgerald01,Doucet05MCsigpro}. For simplicity, in the following we denote the target pdf as $\widetilde{\pi}(\vec{x})$. When direct sampling from $\widetilde{\pi}(\vec{x})$ in the space $\mathbb{R}^L$ is unfeasible, a common approach is the use of {\it Gibbs-type samplers} \citep{Robert04}. This type of methods split the complex sampling problem into simpler univariate cases. Below we briefly summarize some well-known  Gibbs-type algorithms.
\newline
\newline
 {\bf Gibbs sampling.} Let us denote as ${\bf x}^{(0)}$ a randomly chosen starting point. At iteration $k \geq 1$, a Gibbs sampler obtains the $\ell$-th component ($\ell=1, \ldots, L$) of $\vec{x}$, $x_\ell$, drawing from the full conditional $\widetilde{\pi}_{\ell}(x|\vec{x}_{1:\ell-1}^{(k)}, \vec{x}_{\ell+1:L}^{(k-1)})$ given all the information available, namely:
\begin{enumerate}
\item Draw $x_{\ell}^{(k)} \sim \widetilde{\pi}_{\ell}(x|\vec{x}_{1:\ell-1}^{(k)}, \vec{x}_{\ell+1:L}^{(k-1)})$ for $\ell=1,\ldots,L$.
\item Set ${\bf x}^{(k)}=[x_{1}^{(k)},\ldots,x_{L}^{(k)}]^{\top}$.
\end{enumerate}
The steps above are repeated for $k=1,\ldots, N_G$, where $N_G$ is the total number of Gibbs iterations. However, even sampling from $\widetilde{\pi}_{\ell}$ can often be complicated.
In these cases, another efficient Monte Carlo technique (e.g., RS or the MH algorithm) must be employed within the Gibbs sampler. Several alternatives have been proposed for sampling efficiently from the full-conditional pdfs  \citep{Gilks92,Gorur08rev,Hoermann95,koch2007gibbs,Shao13,Gilks95,Meyer08,MartinoA2RMS}.
\newline
\newline
 {\bf  Hit and Run.}  The Gibbs sampler only allows movements along the axes. In certain scenarios, e.g., when the variables $x_\ell$ are highly correlated, this can be an important limitation that slows down the convergence of the chain to the stationary distribution. The {\it Hit and Run sampler} is a valid alternative. Starting from ${\bf x}^{(0)}$, at the $k$-th iteration, it applies the following steps:
 \begin{enumerate}
 \item Choose uniformly a direction ${\bf d}^{(k)}$ in $\mathbb{R}^L$. For instance, it can be done drawing $L$ samples $v_\ell$ from a standard Gaussian $\mathcal{N}(0,1)$, and setting 
 $$
 {\bf d}^{(k)}=\frac{{\bf v}}{\sqrt{{\bf v}{\bf v}^{\top}} },
 $$   
 where ${\bf v}= [v_1,\ldots, v_L]$.
\item Set ${\bf x}^{(k)} ={\bf x}^{(k-1)}+\lambda^{(k)} {\bf d}^{(k)}$ where  $\lambda^{(k)}$ is drawn from the univariate pdf
$$
p(\lambda)\propto  \widetilde{\pi}({\bf x}^{(k-1)}+\lambda {\bf d}^{(k)}),
$$
where $\widetilde{\pi}({\bf x}_{\ell}^{(k-1)}+\lambda {\bf d}^{(k)})$ is a slice of the target pdf along the direction ${\bf d}^{(k)}$.
%$x_{\ell}^{(k)} \sim \widetilde{\pi}_{\ell}(x|\vec{x}_{1:\ell-1}^{(k)}, \vec{x}_{\ell+1:L}^{(k-1)})$.
\end{enumerate}
Also in this case, we need to able to draw from the univariate pdf $p(\lambda)$ using either some direct sampling technique or another Monte Carlo method (e.g., see \citep{Zhang16}).
%where $\pi_{\ell}(x_\ell|\mathbf{x}_{\neg\ell,k-1})$,  for $\ell=1,\ldots,L$, are the (univariate) full-conditional densities.
%\newline
%\newline
There are several methods similar to the Hit and Run where drawing from a univariate pdf is required; for instance, the most popular one is the {\it Adaptive Direction Sampling} \citep{Gilks94}. 
%the direction is chosen in a different way taking into account the previous generated samples.
\newline
\newline
Sampling from univariate pdfs is also required inside other types of MCMC methods. For instance, this is the case of  {\it exchange-type MCMC} algorithms \citep{MurrayExchange} for handling models with intractable partition functions. In this case, efficient techniques for generating artificial observations are needed. Techniques which generalizes the ARS method, using non-parametric proposals, have been applied for this purpose (see \citep{Rohde14}).

%%%%%%%%%%%%%%%%%%%%%%%%%
\section{Numerical Simulations}
\label{Sect_Simul}
%%%%%%%%%%%%%%%%%%%%%%%%%

In this section we provide different numerical results comparing the AISM methods with several benchmark MCMC techniques such as the ARMS technique \cite{Gilks95}, the adaptive MH method in \cite{Haario01} and the slice sampling \cite{Nea03}.\footnote{An example of preliminary Matlab code of AISM, with the constructions described in Section \ref{sec:logDomain} and the update control rule R3, is provided at \url{https://www.mathworks.com/matlabcentral/fileexchange/54701-adaptive-independent-sticky-metropolis--aism--algorithm}.} 
%
%First, we consider univariate target pdfs in Section \ref{Sect_Simul_0} and then we analyze the multivariate case in Section \ref{Sect_Simul_2}.
%
%n this case, we test both the AISM-within-Gibbs (Section \ref{ASM_within_Gibbs_SIMU}) and stand-alone multivariate AISM approaches (Sections \ref{PiecewiseConst_Simu} and \ref{AdaptiveMix_Simu}).}

%%%%%%%%%%%%%%%%%%%%%%%%%%%
%\subsection{Univariate target distributions}
%\label{Sect_Simul_0}
%%%%%%%%%%%%%%%%%%%%%%%%%%%

%%%%%%%%%%%%%%%%%%%%%%%%%%%
\subsection{Multimodal target distribution}
\label{GMEX}
%%%%%%%%%%%%%%%%%%%%%%%%%%%

We study the ability of different algorithms to simulate multimodal densities (which are clearly non-log-concave).
As an example, we consider a mixture of Gaussians as target density, 
\begin{equation*}
	\widetilde{\pi}(x) = 0.5\mathcal{N}(x;7,1)+0.5\mathcal{N}(x;-7,0.1),
\end{equation*}
where $\mathcal{N}(x;\mu,\sigma^{2})$ denotes the normal distribution with mean $\mu$ and variance $\sigma^{2}$.
The two modes are so separated that ordinary MCMC methods fail to visit one of the modes, or remains indefinitely trapped in one of them.
The goal is to approximate the expected value of the target ($E[X]=0$  with $X\sim \widetilde{\pi}(x)$) via Monte Carlo.
We test the  ARMS method \cite{Gilks95} and the proposed AISM and AISMTM algorithms. For AISM and AISMTM, we consider different construction procedures for the proposal pdf:
\begin{itemize}
\item  P1: the construction given in \citep{Gilks95} formed by exponential pieces, specifically designed for ARMS.
\item  P2: alternative construction formed by exponential pieces obtained by a linear interpolation in the log-pdf domain, given in \citep{MartinoA2RMS}.
\item  P3: the construction using uniform pieces in Eq. \eqref{EqSimpler2}.
\item  P4: the construction using linear pieces in Eq \eqref{EqSimpler3}.
\end{itemize}
Furthermore, for AISM and AISMTM, we consider the Update Rule 2 (R2)  with different parameter $\varepsilon$ and the Update Rule 3 (R3) for the inclusion of a new node in the set $\mathcal{S}_t$  (see Section \ref{AlternSect}).
More specifically, we first test AISM and AISMTM with all the construction procedures P1, P2, P3, and P4 jointly with the rule R3. Then, we test AISM with the construction P4 and the update test R2 with  $\varepsilon\in\{0.005,0.01,0.1,0.2\}$. All the algorithms start with $\mathcal{S}_0=\{-10,-8,5,10\}$ and initial state $x_0=-6.6$. For AISMTM, we have set $M\in\{10, 50\}$.
For each independent run, we perform $T=5000$ iterations of the chain.

The results given in Table \ref{resultsASMTM} are the averages over $2000$ runs, without removing any sample to account for the initial burn-in period. Table \ref{resultsASMTM} shows the Mean Square Error (MSE) in the estimation $E[X]$, the  auto-correlation function $\rho(\tau)$ at different lags, $\tau\in\{1,10,50\}$ (normalized, i.e., $\rho(0)=1$), the approximated Effective Sample Size (ESS) of the produced chain \cite[Chapter 4]{Gamerman97bo}
\begin{equation}
ESS\approx\frac{T}{1+2\sum_{\tau=1}^\infty \rho(\tau)},
\end{equation}
(clearly, $ESS\leq T$), the final number of support points $m_T$ and the computing time normalized with respect to the time spent by ARMS \cite{Gilks95}. For simplicity, in Table  \ref{resultsASMTM}, we have report only the case of R2  with  $\varepsilon\in\{0.005,0.01\}$ however other results are shown in Figure \ref{figTest1}.

AISM and AIMTM outperforms ARMS, providing a smaller MSE and correlation (both close to zero). This is due to ARMS does not allow a complete adaptation of the proposal pdf as highlighted in \citep{MartinoA2RMS}.
The adaptation in AISM and AIMTM provides a better approximation of the target than ARMS, as also indicated  by the  ESS which is substantially higher in the proposed methods.
ARMS is in general slower than AISM for two main reasons. Firstly, the construction P1 (used by ARMS) is more costly since requires the computation of several intersection points \citep{Gilks95}. It is not required for the procedures P2, P3 and P4. Secondly, the effective number of iterations in ARMS is higher than $T=5000$ (the averaged value is $\approx 5057.83$) due to the discarded samples in the rejection step (in this case, the chain is not moved forward). 

Figures \ref{figMSE1}(a)-(b)-(c)-(d) depict the averaged autocorrelation function $\rho(\tau)$ for $\tau=1,\ldots,100$ for the different techniques and constructions. Figures \ref{figMSE1}(e)-(f)-(g)-(g) the Averaged Acceptance Probability (AAP; the value of $\alpha$ of the MH-type techniques) of accepting a new state as function of the iterations $t$. We can see that, with AISM and AIMTM, AAP approaches 1 since $q_t$ becomes closer and closer to $\pi$. Figure \ref{figMSE2} shows the evolutions of the number of support points, $m_t$, as function of $t=1,\ldots,T=5000$, again for the different techniques and constructions. 
Note that, with AIMTM and P3-P4, AAP approaches 1 so quickly and  the correlation is so small (virtually zero) that it is difficult to recognize the corresponding curves which are almost constant close to one or zero, respectively. The constructions P3 and P4 provide the better results. In this experiment, P4 seems to provide the best compromise between performance and computational cost. We also test AISM with update R2 for different values of $\varepsilon$ (and different constructions). The number of nodes $m_t$ and AAP as function of $t$ for these cases are shown in Figures \ref{figTest1}. These figures and the results given in Table \ref{resultsASMTM} show that AISM-P4-R2 provides extremely good performance with a small computational cost (e.g, the final number of points is only $m_T\approx 43$ with $\epsilon=0.005$). This proves that the update rule R2 is a very promising choice given the obtained results.

\subsection{Sticky MCMC methods within Gibbs sampling}
\label{ASM_within_Gibbs_SIMU}
%%%%%%%%%%%%%%%%%%%%%%%%%%%

In this example we show that, even in a simple bivariate scenario, AISM schemes can be useful within a Gibbs sampler.
Let us consider the bimodal target density
\begin{equation*}
	\widetilde{\pi}(x_1,x_2)\propto \exp\left(-\frac{(x^2-A+By)^2}{4}-\frac{x^2}{2\sigma_1^2}-\frac{y^2}{2\sigma_2^2}\right),
\end{equation*} 
with $A=16$, $B=10^{-2}$, and $\sigma_1^2=\sigma_2^2=\frac{10^4}{2}$.
Densities with this non-linear analytic form have been used in the literature (cf. \citep{Haario01}) to compare the performance of different Monte Carlo algorithms.
We apply $K$ steps of a Gibbs sampler to draw from $\widetilde{\pi}(x_1,x_2)$, using ARMS \citep{Gilks95}, AISM-P4-R3 and AISMTM-P4-R3  within of the Gibbs sampler to generate samples from the full-conditionals, starting always with the initial support set 
%$$
%\mathcal{S}_0=\{-10, -6, -3.8, 0, 4.3, 7, 10\}.
%$$
$
\mathcal{S}_0=\{-10, -6, -4.3, 0, 3.2, 3.8, 4.3, 7, 10\}$.
From each full-conditional pdf, we draw $T$ samples and take the last one as the output from the Gibbs sampler.
We also apply a standard MH algorithm with a random walk proposal $q(x_{\ell,t}|x_{\ell,t-1}) \propto \exp((x_{\ell,t}-x_{\ell,t-1})^2/(2\sigma_p^2))$ for $\ell \in \{1,2\}$, $\sigma_p \in \{1,2,10\}$, $1 \le t \le T$. Furthermore, we test an adaptive parametric approach (as suggested in \citep{RobRos09}). Specifically, we apply the adaptive MH method in \citep{Haario01}  where the scale parameter of of $q(x_{\ell,t}|x_{\ell,t-1})$ is adapted online, i.e., $\sigma_{p,t}$ varies with $t$ (we set $\sigma_{p,0}=3$). Finally, we consider the application of the slice sampler \citep{Nea03}. For both the standard MH and the slice samplers, we have used the function \texttt{mhsample.m} and  \texttt{slicesample.m} directly provided  by MATLAB (a preliminary code of AISM is also available at \texttt{Matlab-FileExchange} webpage).

%
%{\color{red} Let us denote as $x_{i,n}^{(\ell)}$ the value of the $i$-th coordinate (with $i \in \{1,2\}$) at the $\ell$-th iteration of the Gibbs sampler and the $n$-th iteration of the internal MCMC.
%
We consider two initializations for all the methods-within-Gibbs: ({\bf In1}) $x_{\ell,0}^{(k)}=1$; ({\bf In2}) $x_{\ell,0}^{(k)}=1$ and $x_{\ell,0}^{(k)}=x_{\ell,T}^{(k-1)}$ for $k=1,\ldots,N_G$.
We uses all the samples to estimate four statistics that involve the first four moments of the target: mean, variance, skewness and kurtosis. 
%
%Fig. \ref{fig:Target}(a) illustrates the target $\widetilde{\pi}({\bf x})$, and Figs. \ref{fig:Target}(b)--(c) show the {\it mean absolute error} (MAE) as a function of $T$ for the different techniques averaged over $1000$ runs.
%
Table \ref{TableToyExGibbs} provides the mean absolute error (MAE; averaged over 500 independent runs) for each of the four statistics estimated, and the time required by the Gibbs sampler (normalized by considering $1.0$ to be the time required by ARMS with $T=50$).

%\vspace{-0.2 cm}

First of all, we notice that AISM outperforms ARMS and the slice sampler for all values of $T$ and $N_G$,  in terms of performance and computational time.
%showing  that the AISM adaptive structure speeds up the convergence of the Markov chain.
%
%For instance, AISM with only $T=3$ provides better results than ARMS with $T=50$, saving $95\%$ of the computation time.
%
Regarding the use of the MH algorithm within Gibbs, the results depend largely on the choice of the variance of the proposal, $\sigma_p^2$, and the initialization, showing the need for adaptive MCMC strategies. For a fixed value of $T \times N_G$, the AISM schemes provide results close to the smallest  averaged MAE for {\bf In1} and the best results for {\bf In2}  with a slight increase in the computing time, w.r.t. the standard MH algorithm. Finally, Table \ref{TableToyExGibbs} shows the advantage of the non-parametric adaptive independent sticky approach w.r.t. the parametric adaptive approach \citep{RobRos09,Haario01}.

%%%%%%%%%%%%%%%%%%%%%%%%%%%%%%
\section{Conclusions}
\label{Sect_Concl}
%%%%%%%%%%%%%%%%%%%%%%%%%%%%%%
In this work, we have introduced a new class of adaptive MCMC algorithms for all-purposes stochastic simulation. We have discussed the general features of the novel family, describing the different parts which form a generic sticky adaptive MCMC algorithm. The proposal density used in the new class is  adapted on-line, constructed employing non-parametric procedures. The name ``sticky''  remarks that the  proposal pdf approaches progressively more and more the target. Namely,  a complete adaptation of the shape of the proposal is obtained (unlike when a parametric proposal is used).  The role of the update control test for the inclusion of new support points has been investigated. The designed of this test is extremely important since controls the trade-off between computational cost and the efficiency of the resulting algorithm. Moreover, we have discussed how the combined design of a suitable proposal construction and a proper update test ensures the ergodicity of the generated chain.

Two specific sticky schemes, AISM and ASMTM, have been proposed and tested exhaustively in different numerical simulations.  The numerical results show the efficiency of the proposed algorithms with respect other different benchmark adaptive MCMC methods. Furthermore, we have showed that other algorithms already introduced in the literature are encompassed within the novel class of methods. A detailed description of the related works in the literature and their range of applicability are also provided, which is particularly useful for the interested practitioners and researchers. 
%as well as a discussion about the range of applicability of the novel class of techniques. 
The novel methods can be applied both as a stand-alone algorithm or within any Monte Carlo approach that requires sampling from univariate densities (e.g., the Gibbs sampler, the hit-and-run algorithm or adaptive direction sampling). A promising future line is designing suitable constructions of the proposal density in order to allow the direct sampling from multivariate target distributions (similarly as  \citep{Hoermann95b,Hormann03,Karawatzki06,Leydold98,Leydold98b}). However, we remark that the structure of the novel class of methods is valid regardless of the dimension of the target.% but the challenge is finding efficient procedures to build the proposal pdfs in high-dimensional spaces.

%%% future lines??

 % Different interpolation strategies for the construction of the adaptive nonparametric densities are discussed. %We have been able to prove the ergodicity of the AISMTM algorithm, thus extending previous results in the literature and using conditions which are automatically satisfied by our proposal densities.
%We found that the performance of the ARMS depend crucially on the choice of the initial support points, whereas our AISMTM is robust with respect to this choice. 
% Moreover, the multiple-mode and heavy-tail target examples show that the AISMTM, as opposed to the ARMS, is efficient in exploring the sample space. 
 %The simulation experiments show that the proposal construction methods with uniform pieces and the one with linear pieces in the density domain are the most efficient
 
%{\color{red} future lines: completare il multi dim construction, e sempre per il multi dim....disegnare schemi in cui si possano aggiungere o togliere i punti to control the comp cost ....$\epsilon$ adptive or as function of the derivative... }

%%%%%%%%%%%%%%%%%%%
\section{Acknowledgments}
%%%%%%%%%%%%%%%%%%%
This work has been supported by DISSECT (TEC2012-38058-C03-01), by the BBVA Foundation through project MG-FIAR (ÒI Convocatoria de Ayudas Fundaci—n BBVA a Investigadores, Innovadores y Creadores CulturalesÓ), by the Italian Ministry of Education, University and Research (MIUR), by PRIN 2010-11 grant, by the European Union (Seventh Framework Programme FP7/2007-2013) under grant agreement no:630677. %The authors would also like to thank Joaqu\'in M\'iguez (Universidad Carlos III de Madrid) for many useful comments on a preliminary version of this paper.

%\newpage

%%%%%%%%%%%%%%%%%
%%%%%%%%%%%%%%%%%
%\bibliographystyle{plainnat}
%\bibliographystyle{apalike}
%\bibliographystyle{ims}

%%%%%%%%%%%%%%%%%
%%%%%%%%%%%%%%%%%

\appendix

%{\color{red}
%%%%%%%%%%%%%%%%%%%%%%%%%%%%%%%%%%%%%%%%%%%%
\section{Ergodicity of the generated chain}
\label{TeoApp}
%%%%%%%%%%%%%%%%%%%%%%%%%%%%%%%%%%%%%%%%%%%%
%}

The updated test in the sticky methods considers the variable $z$, which is always different from the current state $x_{t}$.
%
%This approach leads to a proposal, $\pi_t(x)$, which is independent of the current state of the chain, $x_n$.
%
Thus, the proposal is independent from the current state and the convergence of the resulting chain to the stationary (bounded) target density is ensured by Theorem 2 in \citep{Holden09}. Indeed, the sticky MCMC algorithms also satisfy the {\it strong Doeblin condition} as required in \citep{Holden09}. Namely, this condition is fulfilled if, given a proposal pdf $\widetilde{q}_t(x|\mathcal{S}_t)$, there exists some $a_{t}\in(0,1]$, such that  
\begin{equation}
	\frac{1}{a_t}\widetilde{q}_t(x|\mathcal{S}_t)\geq \widetilde{\pi}(x) \qquad \forall x \in \mathcal{X}.
\label{Doeblin}
\end{equation}
Hence, we need to ensure the existence of a suitable value $a_{t}\in(0,1]$, for all $t\in \mathbb{N}$. 
Denoting $\widetilde{q}_t(x|\mathcal{S}_t) = \frac{1}{c_t} q_t(x|\mathcal{S}_t)$ and $\widetilde{\pi}(x) = \frac{1}{c_\pi} \pi(x)$, Eq. \eqref{Doeblin} can be rewritten as
\begin{equation}
	a_t \le \frac{c_\pi}{c_{t}} \frac{q_t(x|\mathcal{S}_t)}{\pi(x)} \qquad \forall x \in \mathcal{X}.
\label{Doeblin2}
\end{equation}
Moreover, since  $\min\{1,x\} \le x$, in order to fulfill Eq, \eqref{Doeblin2},  a possible value $a_{t}$ is
\begin{equation}
\label{a_t}
	a_t = \min\left\{1,\frac{c_\pi}{c_{t}} \min_{x \in \mathcal{X}}\left\{\frac{q_t(x|\mathcal{S}_t)}{\pi(x)}\right\}\right\}.
\end{equation}
Furthermore, we can always guarantee that $q_t(x|\mathcal{S}_t) > \pi(x)$ in the tails  by using an appropriate construction of the tails of the proposal (exponential tails or  heavier tails, as described in \citep{MartinoA2RMS}).
Thus, we can use the $a_t$ in Eq. \eqref{a_t} for $ x\in \mathcal{I}$, where  $\mathcal{I} = \cup_{i=1}^{m_t}{\mathcal{I}_i}=[s_1,s_{m_t}]$ and $a_t=1$ for $x\notin \mathcal{I}$ since $q_t(x|\mathcal{S}_t) > \pi(x)$ in the tails. For the constructions considered in this work, the value $a_t$ in Eq. \eqref{a_t} satisfies all the conditions required in \citep{Holden09}: $a_t \in(0, 1]$ and $a_t \to 1$ as $t \to \infty$, since $q_t(x|\mathcal{S}_t) \to \pi(x)$ as $t \to \infty$ (when the adaptation is not stopped) and thus also $c_{t} \to c_\pi$ as $t \to \infty$. Therefore, we have also that
\begin{equation*}
	\lim_{T \to \infty}{\prod_{t=1}^{T}{(1-a_t)}} \to 0,
\end{equation*}
as required in Theorem 2 of \citep{Holden09}. When the adaptation is stopped at some iteration $t^*$, due to the chosen update rule, the algorithm becomes a standard MCMC technique satisfying the balance condition \citep{Robert04,MartinoJesse}, so that the ergodicity for $t\geq t^*$ is automatically ensured. For $t< t^*$, the ergodicity is ensured  by Theorem 2 in \citep{Holden09} as described previously.  

\begin{table*}[hbt]
%\begin{sidewaystable}[p]
%\begin{table}[p]
\setlength{\tabcolsep}{2pt}
\def\marginwidth{1.5mm}
\begin{center}
%\label{resultsASM}
%\small
\begin{tabular}{|l@{\hspace{\marginwidth}}|c@{\hspace{\marginwidth}}||c@{\hspace{\marginwidth}}|c@{\hspace{\marginwidth}}|c@{\hspace{\marginwidth}}|c@{\hspace{\marginwidth}}||c@{\hspace{\marginwidth}}|c@{\hspace{\marginwidth}}|}
%\begin{tabularx}{\textwidth}{|c|c|c|c|c|c|c|c|}
\hline
{\bf Algorithm} & {\bf MSE} & ${\bm \rho{\bf(1)}}$ & ${\bm \rho{\bf(10)}}$ & ${\bm \rho{\bf(50)}}$ & {\bf ESS} &${\bm m_{T}} $ &{\bf Time} \\ %& {\bf EI}  \\
\hline
\hline
ARMS \citep{Gilks95}  & 10.04  & 0.4076 &0.3250 &0.2328 & 89.12  &118.19&  1.00 \\% & 5057.83\\     
%\hline                                                                          
%ARMS-2  & 15.6756  & 0.8955 &0.7210 &0.4639 & & 7.6126  &  0.1195& 5003.612\\     
%\hline                                                                          
%ARMS-3  & 0.2398  & 0.8753 &0.4410 &0.0296 & &131.3360 &  0.3589& 5127.336\\     
%\hline                                                                          
%ARMS-4  & 0.2874   & 0.8882 &0.4758 &0.0418 & & 42.8872 &  0.2291& 5038.887 \\    
 \hline                                                                         
  \hline 
AISM-P1-R3   & 3.0277  & 0.1284  &0.1099 &0.0934 & 235.76 & 152.63  &  1.23 \\%& 5000\\         
%\hline                                                                          
AISM-P2-R3   & 2.9952  & 0.1306  &0.1125 &0.0929 & 235.01 & 71.14 &  0.27\\%& 5000\\         
%\hline                                                                          
AISM-P3-R3   & 0.0290  & 0.0535 &0.0165  &0.0077 &  609.05  & 279.65 &  0.65\\%& 5000\\         
%\hline                                                                          
AISM-P4-R3    & 0.0354  & 0.0354 &0.0195   & 0.0086&  608.76 &84.87&  0.33\\%& 5000\\         
\hline
 \hline 
\multirow{1}{*}{AISMTM-P1} ($M=10$)  & 0.6720 & 0.0726 &0.0696 & 0.0624 & 336.84    &159.01  & 2.35 \\%&5000\\
\hspace{0.85cm}R3\hspace{0.85cm}($M=50$) & 0.1666 & 0.0430 &0.0395 & 0.0316 & 617.10 &160.75  & 5.45\\% &5000\\
%\hline
AISMTM-P2 ($M=10$)  & 0.5632 & 0.0588 &0.0525 &0.0443  & 440.23 & 72.16   & 1.13 \\%&5000\\
\hspace{0.85cm}R3\hspace{0.85cm}($M=50$)  & 0.1156 & 0.0345 &0.0303 &0.0231  & 746.45 &72.53   & 4.38\\% &5000\\
%\hline
AISMTM-P3 ($M=10$)  & 0.0105  & 0.0045 &0.0001 &0.0001 & 4468.10 & 315.78  & 2.60 \\%&5000\\
\hspace{0.85cm}R3\hspace{0.85cm}($M=50$)  & 0.0099 & 0.0041  &0.0001 &0.0001 & 4843.81 &360.73 &10.59 \\%&5000\\
%\hline
AISMTM-P4 ($M=10$)  &  0.0108 & 0.0036  &0.0011 &0.0014 & 3678.79 &92.67  & 1.86 \\%&5000\\
\hspace{0.83cm}R3\hspace{0.85cm}($M=50$) &  0.0098 &  0.0001 &0.0001 &0.0001 & 4912.07 &101.78 & 7.25\\% &5000\\
\hline
\hline
\multirow{2}{*}{AISM-P4-R2} ($\varepsilon=0.01$)  & 0.0412  & 0.0407 &0.0213   & 0.0074 &  604.95  &35.01 & 0.11 \\%& 5000\\  
\hspace{2cm}($\varepsilon=0.005$)  &  0.0321 & 0.0360 &0.0181   & 0.0072 &  610.01 & 43.32 & 0.20 \\%& 5000\\  
\hline
%\end{tabularx}
\end{tabular}
\end{center}
\vspace{-0.5cm}
\caption{ {\bf(Ex-Sect-\ref{GMEX})} For each algorithm, the table shows the mean square error (MSE), the autocorrelation ($\rho(\tau)$) at different lags, the effective sample size (ESS), the final number of support points ($m_{T}$), the computing times normalized w.r.t. ARMS (Time).} %and the effective number of iterations (EI; ARMS requires more iterations for obtaining the same number of samples due its rejection sampling steps, where the chain is not moved forward).}}
\label{resultsASMTM}
%\end{sidewaystable} 
\end{table*}
%\end{figure*}
%%%%%%%
%%%%%%%%
\begin{figure*}[hbt]
\begin{center}
\centerline{
%width=3.95cm
%\subfigure[NSP, AISM-1]{\includegraphics[width=140pt, height=88pt, angle=0]{POI_ASM_P1.pdf}}
\subfigure[$m_t$ (AISM-P1)]{\includegraphics[width=125pt]{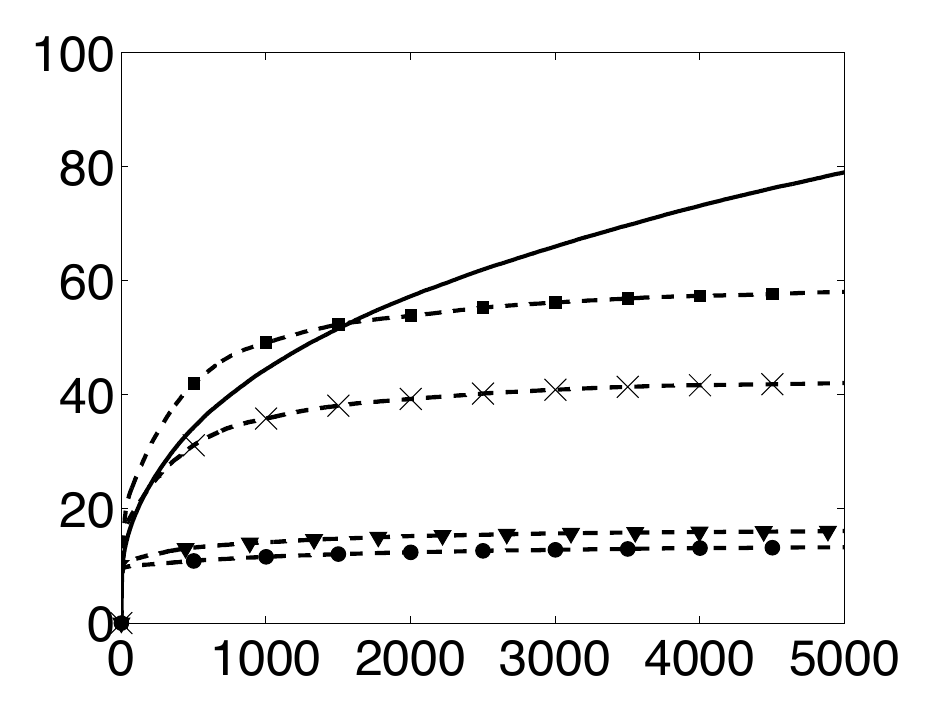}}
\subfigure[$m_t$ (AISM-P2)]{\includegraphics[width=125pt]{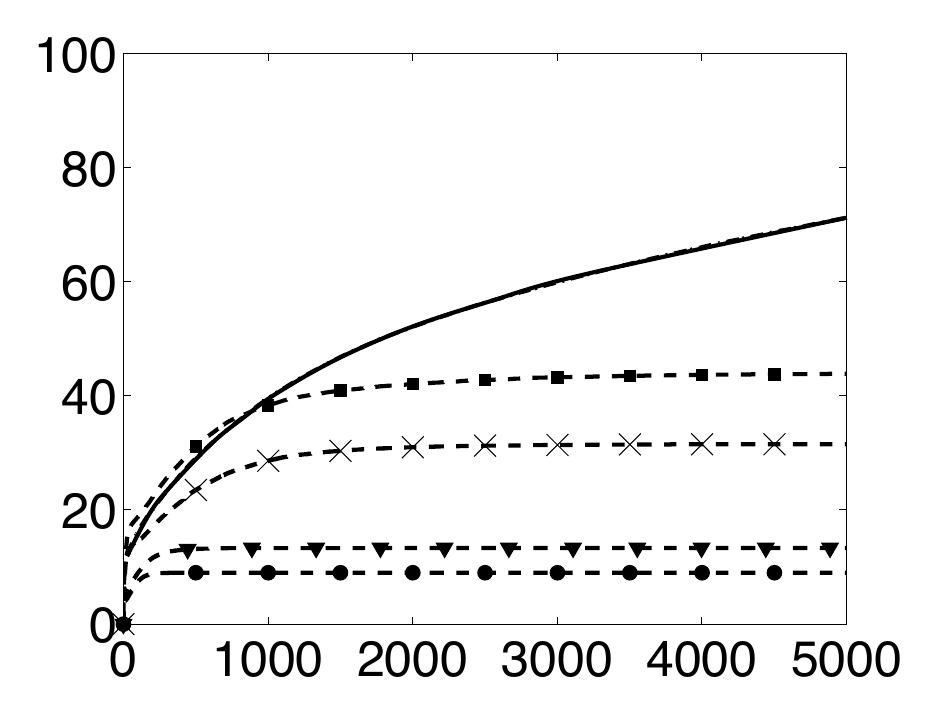}}
\subfigure[$m_t$ (AISM-P3)]{\includegraphics[width=125pt]{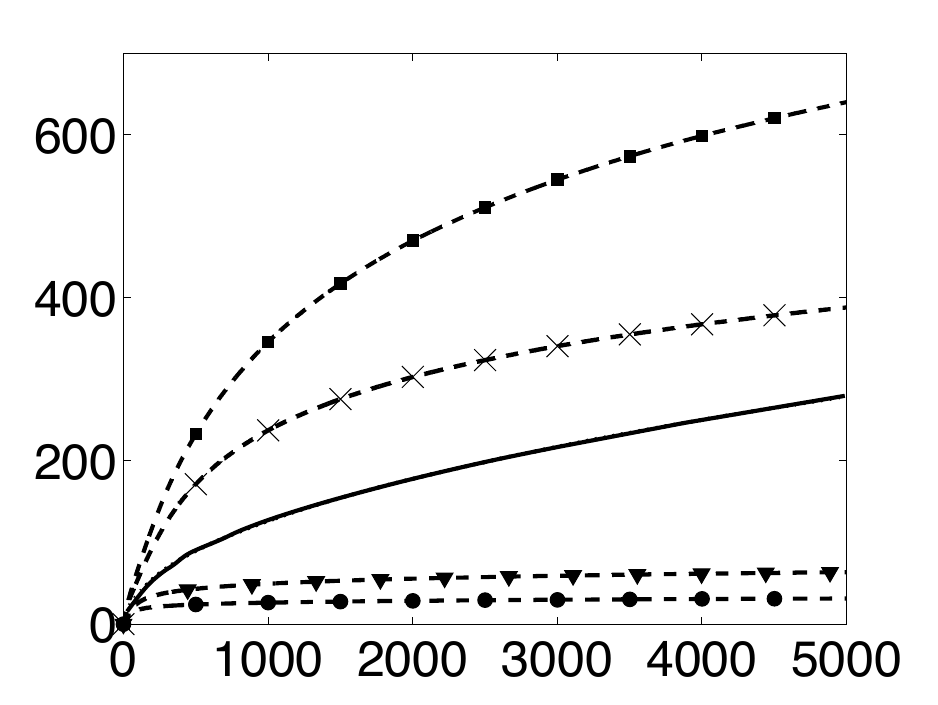}}
\subfigure[$m_t$ (AISM-P4)]{\includegraphics[width=125pt]{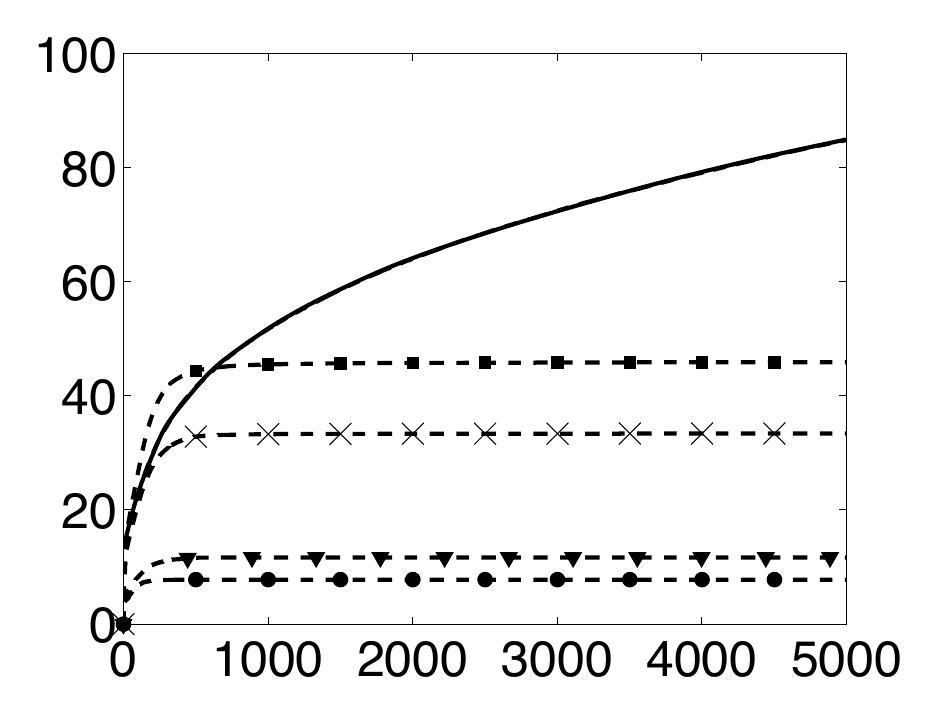}}
}
\centerline{
%width=3.95cm
%\subfigure[NSP, AISM-1]{\includegraphics[width=140pt, height=88pt, angle=0]{POI_ASM_P1.pdf}}
\subfigure[AAP, (AISM-P1)]{\includegraphics[width=125pt]{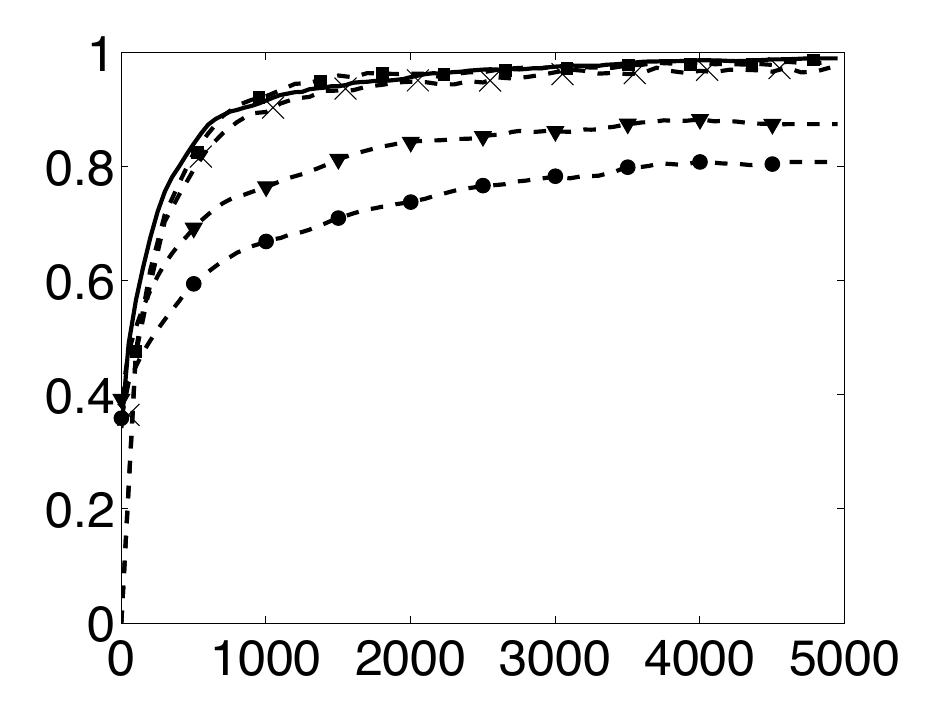}}
\subfigure[AAP, (AISM-P2)]{\includegraphics[width=125pt]{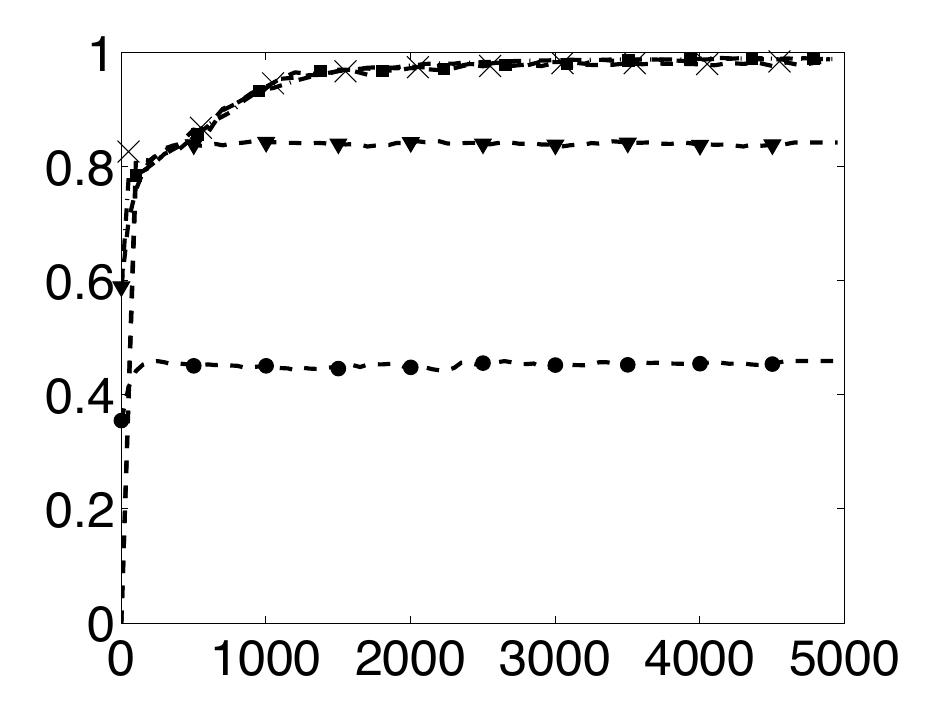}}
\subfigure[AAP, (AISM-P3)]{\includegraphics[width=125pt]{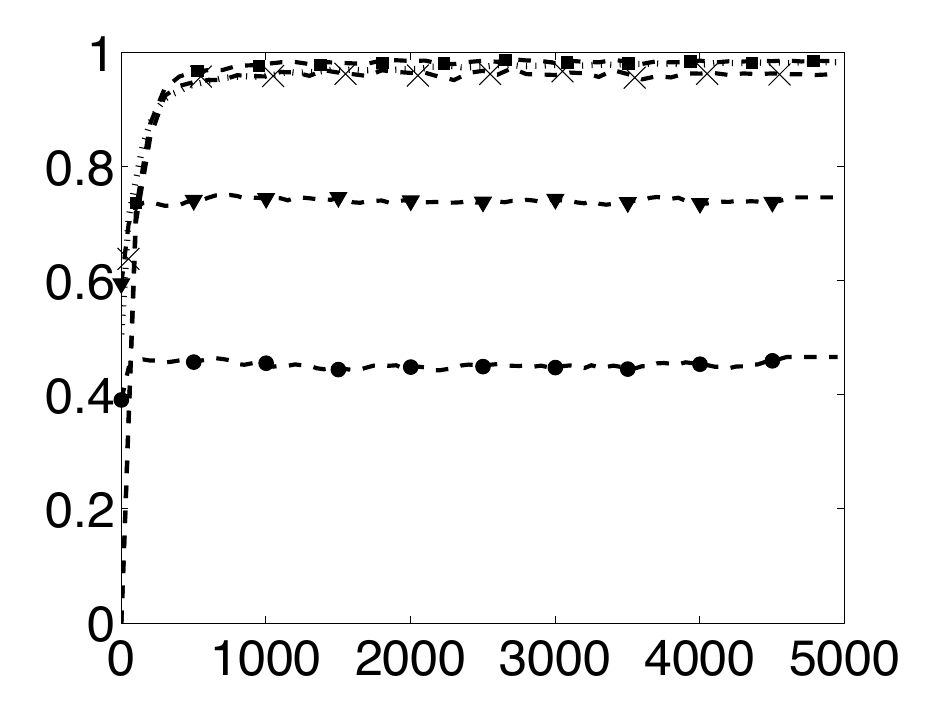}}
\subfigure[AAP, (AISM-P4)]{\includegraphics[width=125pt]{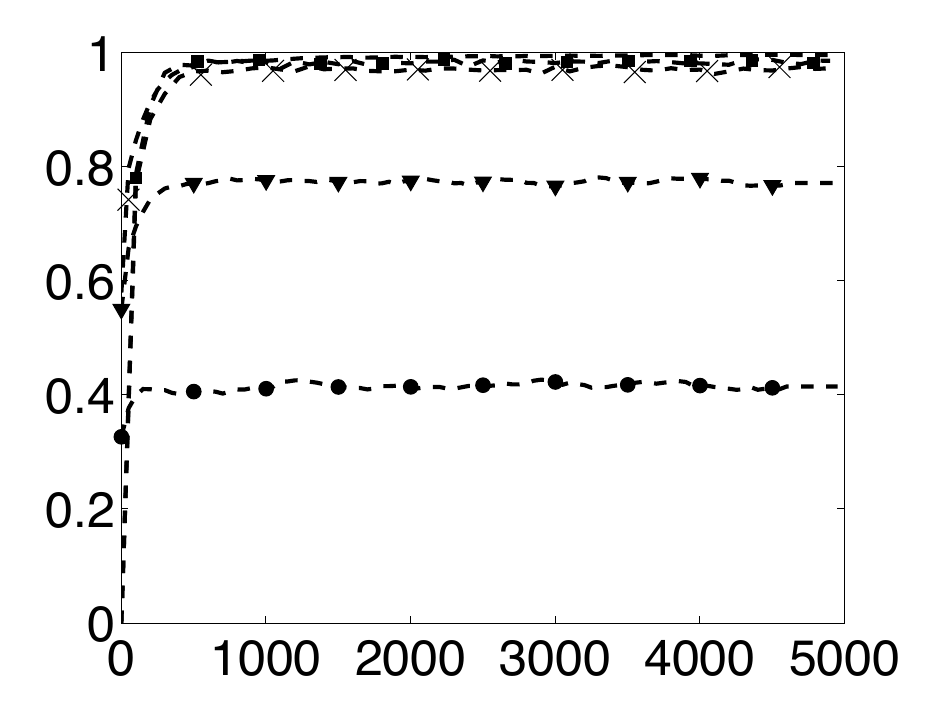}}
}
\end{center}
\vspace{-0.5cm}
\caption{{\bf(Ex-Sect-\ref{GMEX})} Evolution of the number of support points $m_t$  and Averaged Acceptance Probability (AAP), as function of $t=1,\ldots,T$ for AISM, for different constructions,  and update rule R2 with $\varepsilon=0.005$ (square), $\varepsilon=0.01$ (cross), $\varepsilon=0.1$ (triangle) and $\varepsilon=0.2$ (circle). Moreover, in Figures (a)-(b)-(c)-(d) the evolution of $m_t$ of AISM with the update rule R3 is also shown with solid line. Note that the range of values in Figs. (a)-(b)-(c)-(d)  is different.  } 
\label{figTest1}
\end{figure*}

%%%%%%%
%%%%%%%%

\begin{figure*}[p]
\begin{center}
%\centerline{
%\subfigure[MSE, ARMS]{\includegraphics[width=120pt]{MSE_ARMSv2.pdf}}
%\subfigure[MSE, AISM]{\includegraphics[width=120pt]{MSE_ASMv2.pdf}}
%\subfigure[MSE, AISMTM ($M=10$)]{\includegraphics[width=125pt]{MSE_ASMTM_10.pdf}}
%\subfigure[MSE, AISMTM ($M=50$)]{\includegraphics[width=125pt]{MSE_ASMTM_50.pdf}}
%}
\centerline{
\subfigure[ $\rho(\tau)$ (ARMS)]{\includegraphics[width=120pt]{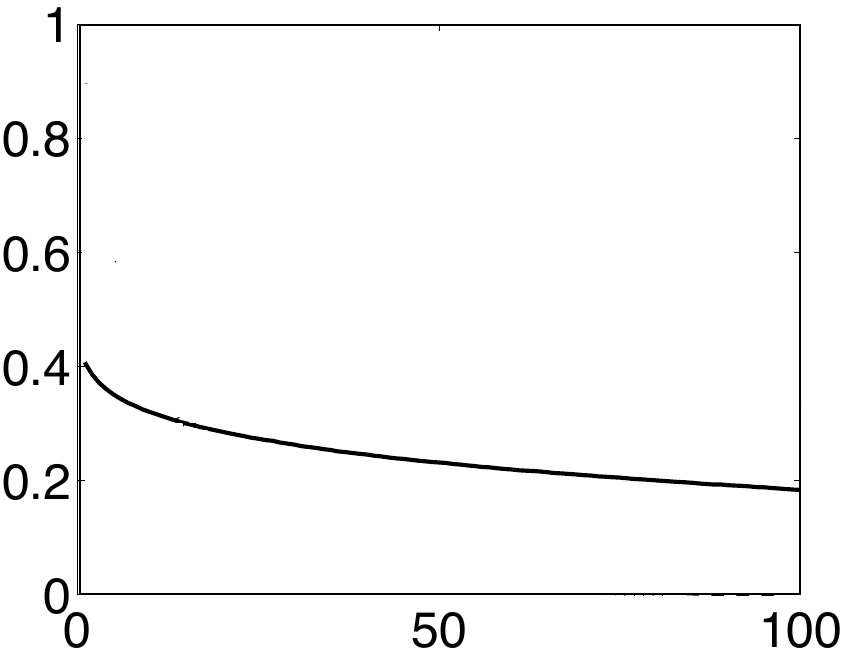}}
\subfigure[$\rho(\tau)$ (AISM)]{\includegraphics[width=125pt]{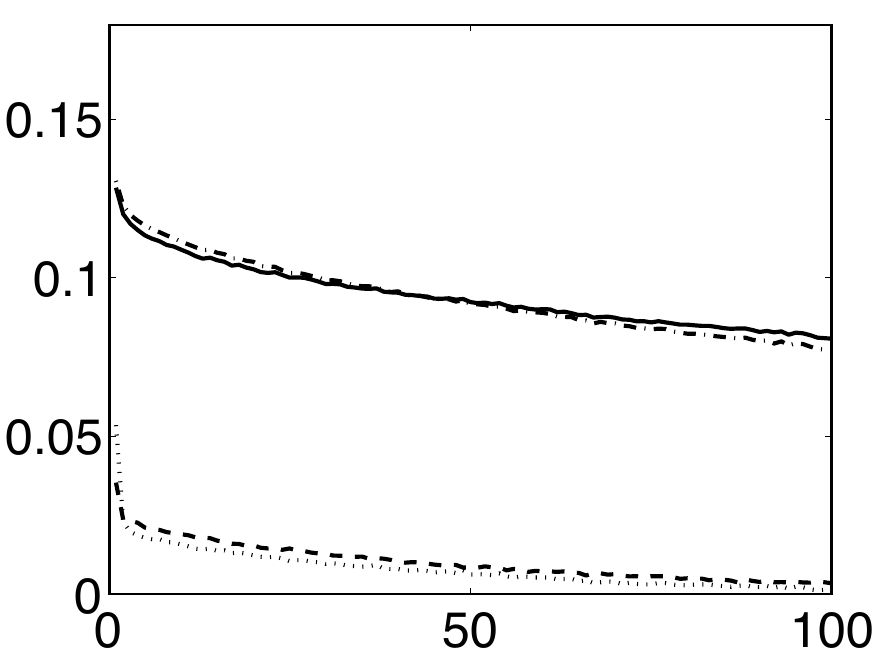}}
\subfigure[$\rho(\tau)$ (AISMTM-$M=10$)]{\includegraphics[width=125pt]{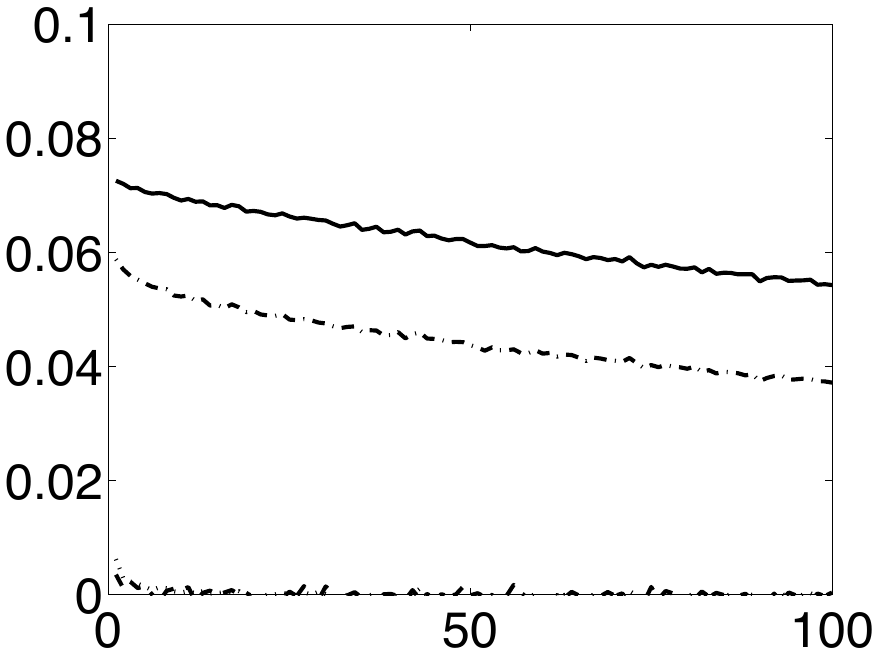}}
\subfigure[$\rho(\tau)$ (AISMTM-$M=50$)]{\includegraphics[width=125pt]{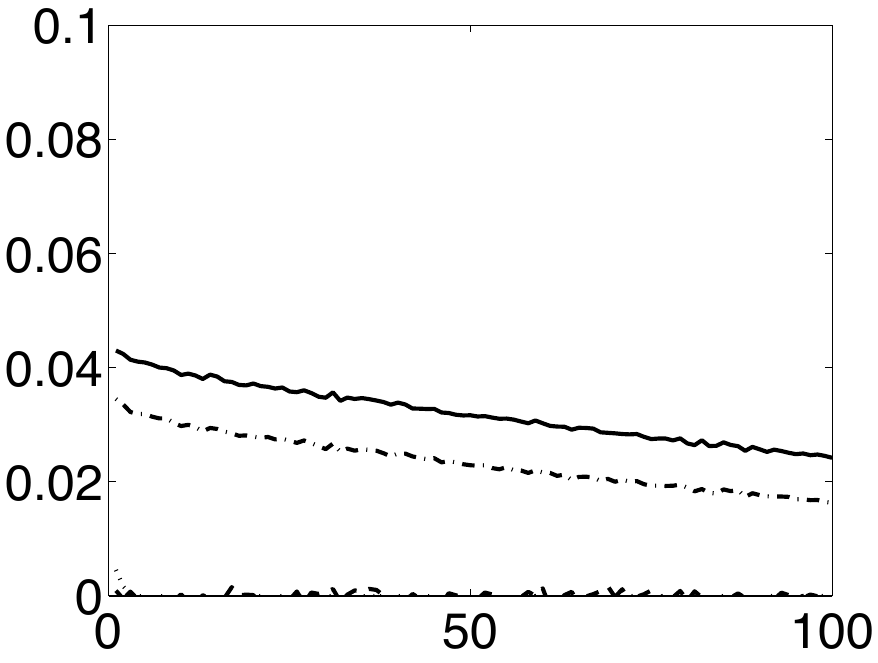}}
}
\centerline{
\hspace{-0.2cm}
\subfigure[AAP (ARMS)]{\includegraphics[width=117pt]{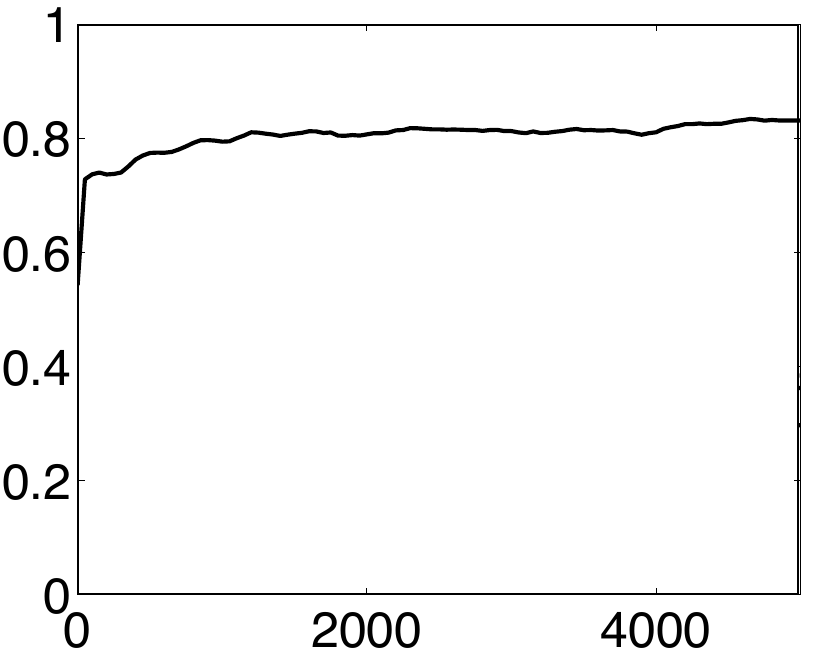}}
\hspace{0.2cm}
\subfigure[AAP (AISM)]{\includegraphics[width=117pt]{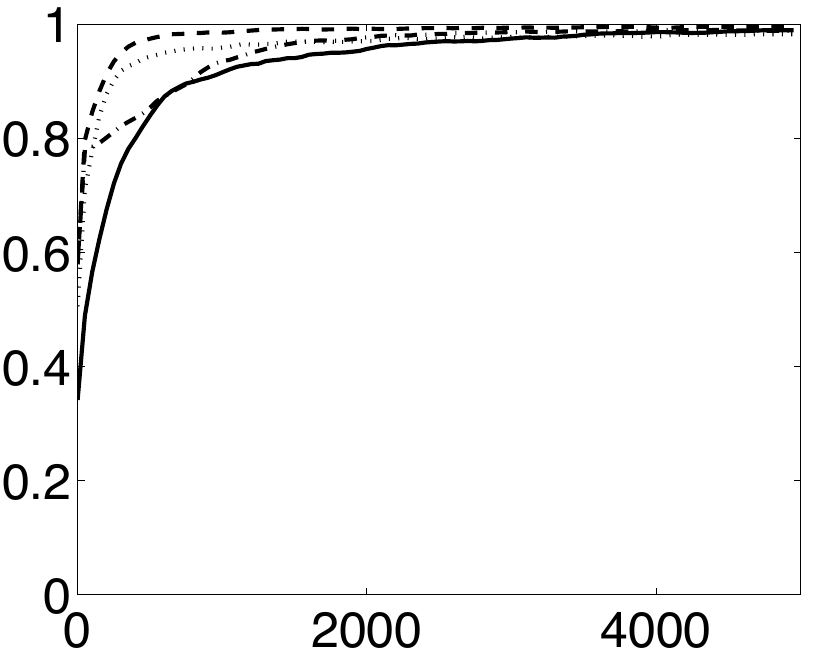}}
\hspace{0.2cm}
\subfigure[AAP (AISMTM-$M=10$)]{\includegraphics[width=117pt]{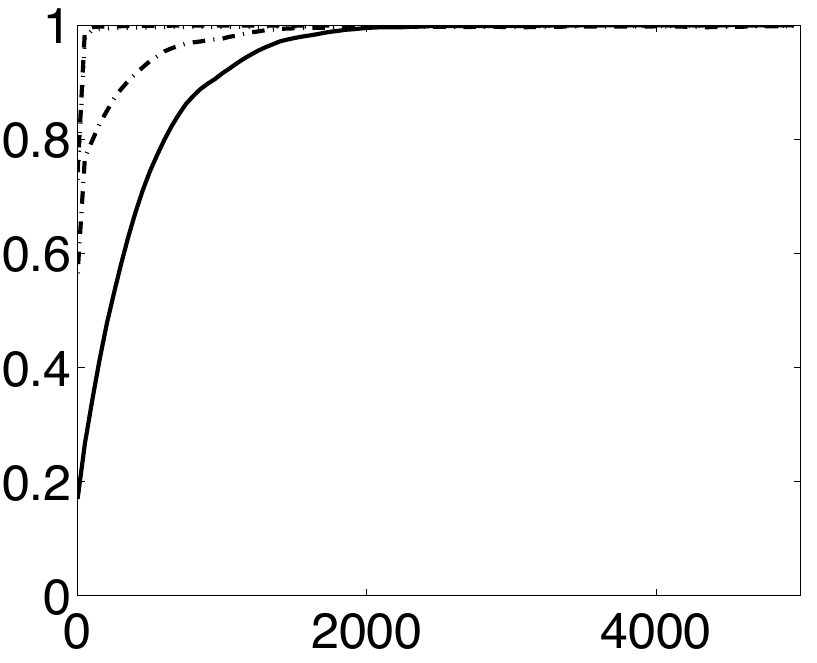}}
\hspace{0.2cm}
\subfigure[AAP (AISMTM-$M=50$)]{\includegraphics[width=117pt]{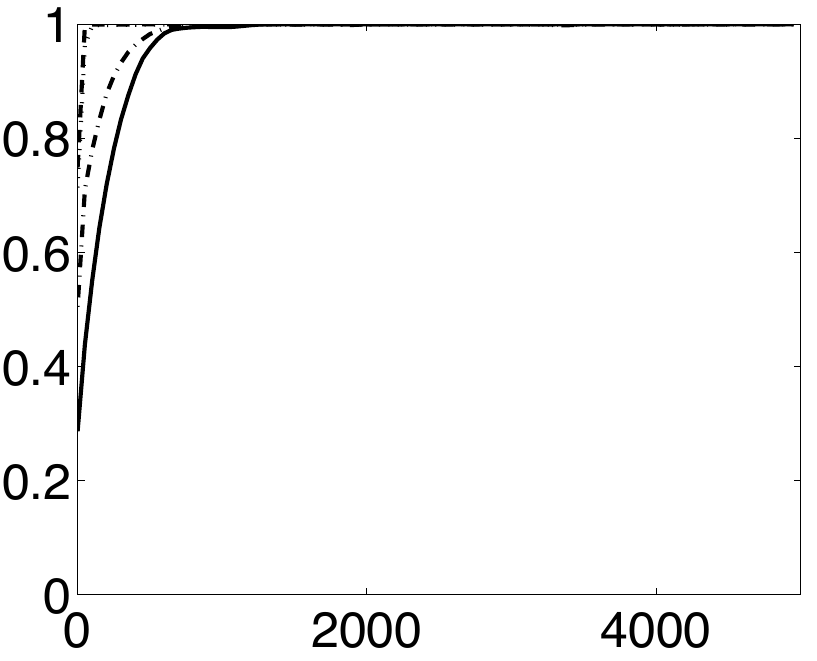}}
}
\end{center}
\vspace{-0.5cm}
\caption{{\bf  (Ex-Sect-\ref{GMEX})}
Autocorrelation Function $\rho(\tau)$ at lags from 1 to 100 and Averaged Acceptance Probability (AAP) as function of $t$, for the different methods. In each plot: P1 (solid line), P2 (dashed-dotted line), P3 (dotted line) and  P4 (dashed line). Note the different range of values of $\rho(\tau)$.}
\label{figMSE1}
\end{figure*}

%%%%%%%%%%%%%
%%%%%%%%%%%%%
%%%%%%%%%%%%%
%%%%%%%%%%%%%
\begin{figure*}[p]
\begin{center}
\centerline{
\subfigure[$m_t$ (ARMS)]{\includegraphics[width=125pt]{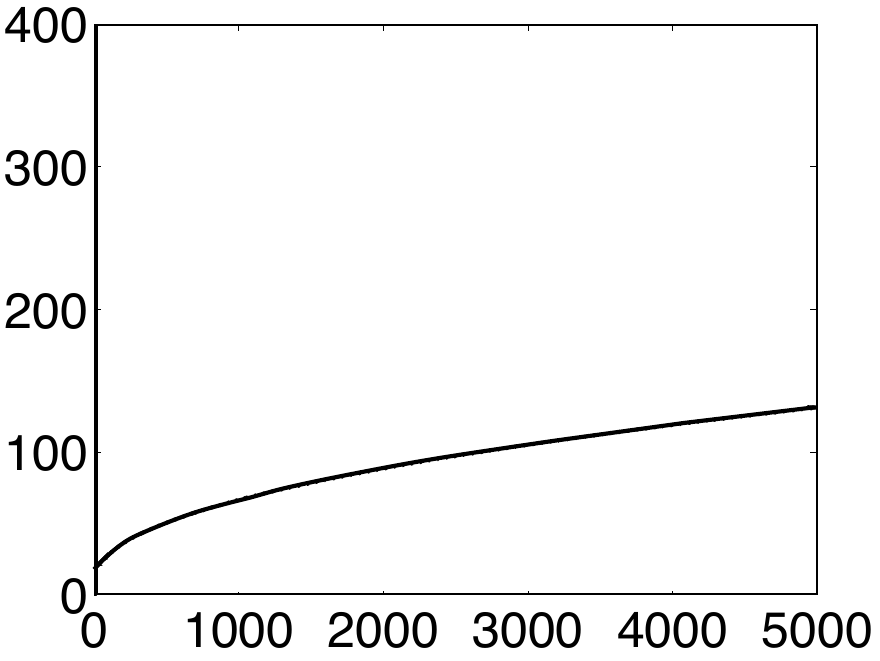}}
\subfigure[$m_t$ (AISM)]{\includegraphics[width=125pt]{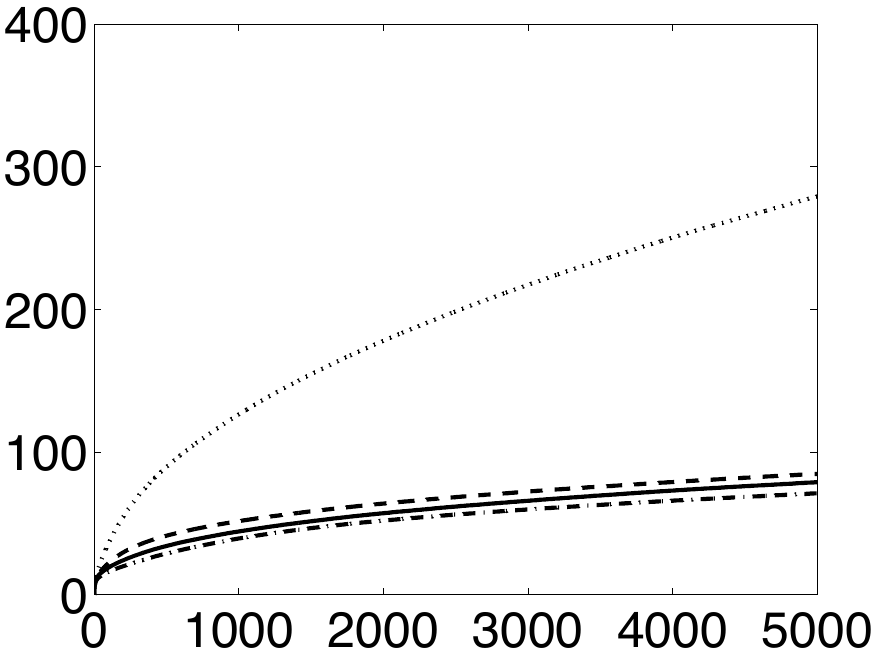}}
\subfigure[$m_t$ (AISMTM-$M=10$)]{\includegraphics[width=125pt]{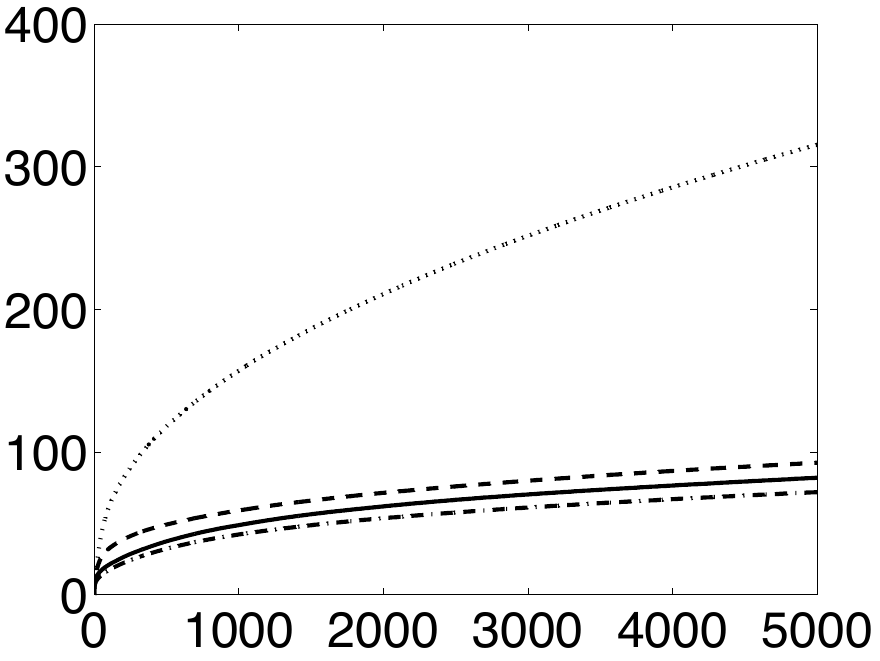}}
\subfigure[$m_t$ (AISMTM-$M=50$)]{\includegraphics[width=125pt]{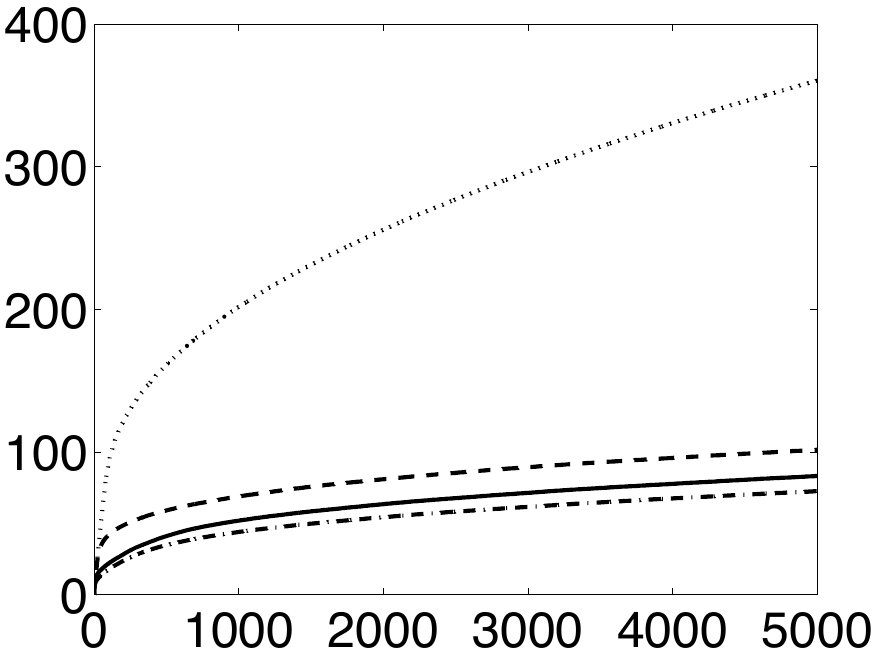}}
}
\end{center}
\vspace{-0.5cm}
\caption{{\bf  (Ex-Sect-\ref{GMEX})} Evolution of the number of support points $m_t$ as function of $t=1,\ldots,T$, for the different methods. In each plot: construction P1 (solid line),  P2 (dashed-dotted line), P3 (dotted line) and P4 (dashed line).}
\label{figMSE2}
\end{figure*}

%%%%%%%%%%%%%%%%%%%%%%%%%%%%
%%%%%%%%%%%%%%%%%%%%%%%%%%%%
%%%%%%%%%%%%%%%%%%%%%%%%%%%%
%%%%%%%%%%%%%%%%%%%%%%%%%%%%

%%%%%%%%%%%%%
%%%%%%%%%%%%%%%
\begin{table*}[!hbt]
\def\marginwidth{1.5mm}
\begin{center}
\vspace{-0.2cm}
\scriptsize
\begin{tabular}{|c@{\hspace{\marginwidth}}|c@{\hspace{\marginwidth}}|c@{\hspace{\marginwidth}}|c@{\hspace{\marginwidth}} |c@{\hspace{\marginwidth}}|c@{\hspace{\marginwidth}}|c@{\hspace{\marginwidth}}|c@{\hspace{\marginwidth}}||c@{\hspace{\marginwidth}}||c@{\hspace{\marginwidth}}|}
\multicolumn{10}{c}{{\bf Panel I}} \\
\hline
\multirow{2}{*}{{\bf Technique}}  &  \multirow{2}{*}{${\bf T}$} &  \multirow{2}{*}{ ${\bf N_G}$} & \multirow{2}{*}{{\bf Init.}}  & \multicolumn{4}{c||}{{\bf MAE}} &  \multirow{2}{*}{{\bf Avg. MAE}}   &  \multirow{2}{*}{{\bf Time}}   \\
\cline{5-8}
 & & & &{\bf Mean} & {\bf Variance}  & {\bf Skewness}  & {\bf Kurtosis} & & \\
%  &   &  & & & \
\hline
\multirow{4}{*}{AISM-P4} & $3$ & \multirow{4}{*}{2000}  & \multirow{4}{*}{In1} & 0.878  & 0.781 & 0.437 & 0.223 & 0.579 &   0.066 \\
\cline{2-2}  %\cline{5-10}  
&  $5$ & & & 0.749 &   0.576 &    0.389 &  0.160 & 0.468  &   0.098\\
\cline{2-2}  %\cline{5-10} 
& $10$ & & & 0.266 &  0.057  &  0.136   & 0.020 & 0.120  & 0.178   \\
\cline{2-2}  %\cline{5-10}
 &$50$ & & & 0.101 & 0.041  &  0.051  & {\bf 0.003 } & 0.049 &  0.741 \\
\hline
\hline  
 \multirow{1}{*}{AISMTM-P4}   & 3 &  \multirow{2}{*}{2000} & \multirow{2}{*}{In1} &  0.251 & 0.056  &  0.128  & 0.017 &  0.113 & 0.202 \\ 
 \cline{2-2}
     ($M=5$)  & 10 & & & 0.096 &  {\bf 0.031} &  0.048 &   {\bf 0.003} & 0.044 & 0.642 \\ 
\hline
\hline
\multirow{4}{*}{ARMS}  & $3$ &  \multirow{4}{*}{2000} & \multirow{4}{*}{In1}& 3.408 &11.580 & 3.384 & 11.572 & 7.486 & 0.077 \\
\cline{2-2}  %\cline{5-10}
  & $5$ & & & 3.151 & 9.839 & 2.650 & 7.079 & 5.679 & 0.116 \\
\cline{2-2}  %\cline{5-10}
 & $10$& && 2.798 & 7.665 & 2.024 & 4.124 & 4.152 & 0.223\\
\cline{2-2}  %\cline{5-10}
  & $50$ & &&1.918 & 3.407 & 1.134 & 1.292 & 1.937 &1.000\\
\hline
\hline
 MH  ($\sigma_p=1$) & \multirow{3}{*}{$100$} &   \multirow{3}{*}{2000} & \multirow{3}{*}{In1} & 3.509 &12.308 & 3.671 & 13.666 & 8.288  & 0.602\\
\cline{1-1}  %\cline{5-10}
 MH  ($\sigma_p=2$) & & & & 1.756 & 3.077 & 0.978 & 0.963 & 1.693 &0.602\\
\cline{1-1}  %\cline{5-10}
 MH ($\sigma_p=10$) & & &&  {\bf 0.075} &  0.037 &  {\bf 0.036} & {\bf 0.002} & {\bf 0.038} & 0.602\\
\hline
\hline
MH  ($\sigma_p=1$) &  \multirow{3}{*}{$1000$}  &  \multirow{3}{*}{2000} & \multirow{3}{*}{In1} & 3.508  & 12.302 & 3.665 & 13.624 & 8.274 & 4.052\\
\cline{1-1}  %\cline{5-10}
MH ($\sigma_p=2$) & && &  1.601 & 2.560 & 0.874 & 0.769 & 1.451 & 4.052\\
\cline{1-1}  %\cline{5-10}
MH ($\sigma_p=10$) & & && {\bf 0.074}  & 0.036 & {\bf 0.036} & {\bf 0.002} & {\bf 0.037} & 4.052\\
\hline
\hline
\multirow{4}{*}{ MH  ($\sigma_p=10$)}  &  \multirow{2}{*}{$1$}  & 2000 & \multirow{4}{*}{In1} & 0.697 & 11.598 & 0.883 & 3.622 & 4.200 & \bf 0.033\\
%\cline{3-3} % \cline{5-10}
% &   & 6000 &  & 0.587 & 10.329 & 0.721 & 3.317  & 3.738 & 0.089 \\
 \cline{3-3}  %\cline{5-10}
  &   & 10000 & & 0.493 & 9.881 & 0.611 & 2.905  & 3.472 & 0.162 \\
  \cline{2-3}  %\cline{5-10}
 & $3$ &    \multirow{2}{*}{$2000$} & &  0.352 & 6.510 & 0.290 & 0.927 & 2.019 & 0.042 \\
%\cline{2-2}  %\cline{5-10}
 %& $5$  &   & & 0.205 & 3.916 & 0.134 & 0.460 & 1.178 & 0.053 \\
\cline{2-2}  %\cline{5-10}
 & $10$ & & & 0.085 & 1.411 & 0.043& 0.160 & 0.424 & 0.081 \\
%\cline{2-2}  %\cline{5-10}
% & $50$ & & & \bf 0.077 & 0.053 & \bf 0.037 & 0.007 & 0.043 & 0.290 \\
\hline
\hline
 \multirow{2}{*}{Adaptive MH} & $100$ & \multirow{2}{*}{2000} & \multirow{2}{*}{In1}  & 0.415   & 0.304  &  0.234  & 0.068 & 0.255  &   0.634  \\
 \cline{2-2}
 & $1000$ &  &   & {\bf 0.075} &  0.038 &  0.037 & {\bf 0.002} & {\bf 0.038} & 4.107  \\
\hline
\hline
 \multirow{3}{*}{Slice} & $3$ & \multirow{3}{*}{2000} & \multirow{3}{*}{In1}  & 0.810 & 1.174 & 0.415 & 0.231 & 0.658 &  0.156  \\
 \cline{2-2}
 & $10$ &  & &   0.607 &   0.372  & 0.306  & 0.096   & 0.345 & 0.463  \\
 \cline{2-2}
 & $50$ &  & & 0.156 &   0.043  &  0.077  &  0.007 &   0.071 & 2.311\\
\hline
\multicolumn{10}{c}{\bf Panel II} \\
\hline
\multirow{2}{*}{{\bf Technique}}  &  \multirow{2}{*}{${\bf T}$} &  \multirow{2}{*}{ ${\bf N_G}$} & \multirow{2}{*}{{\bf Init.}}  & \multicolumn{4}{c||}{{\bf MAE}} &  \multirow{2}{*}{{\bf Avg. MAE}}   &  \multirow{2}{*}{{\bf Time}}   \\
\cline{5-8}
 & & & &{\bf Mean} & {\bf Variance}  & {\bf Skewness}  & {\bf Kurtosis} & & \\
\hline
  \multirow{4}{*}{AISM-P4}  & 3 & \multirow{3}{*}{2000} & \multirow{4}{*}{In2}  & 0.138 &   0.055 &   0.070 &   0.006  &  0.067 &   0.066   \\
\cline{2-2}   
%\hline
 & 5&  & & 0.112  &  0.050  & 0.057   & 0.004 &  0.056 &  0.098   \\ 
\cline{2-2}  
& 10&  &  & 0.093  &  0.045 &   0.046 &   {\bf 0.002} & 0.046 & 0.178  \\ 
\cline{2-3} 
& 3& 10000 &  & 0.095 &    {\bf 0.023} &    0.050 &   {\bf 0.002} & 0.042 &  0.335 \\ 
\hline 
\hline 
 \multirow{1}{*}{AISMTM-P4}   & \multirow{3}{*}{3}  &  2000 & \multirow{3}{*}{In2}  & 0.085  &  0.036   & 0.043  &  {\bf 0.002} &  0.042 & 0.202\\ 
  \cline{3-3}
     ($M=5$)  &  & 4000 & & 0.083  &  {\bf 0.028} &  0.042   & {\bf 0.002}   & {\bf0.038}& 0.400 \\ 
      \cline{1-1}   \cline{3-3}
      ($M=10$)  &  & 2000 & & {\bf 0.073}  & 0.031 & {\bf 0.036} &  {\bf 0.002}  & {\bf 0.035}  & 0.316 \\ 
\hline
\hline
\multirow{6}{*}{MH ($\sigma_p=10$)}  & \multirow{3}{*}{$1$}  & 10000  &   \multirow{6}{*}{In2} & 0.178 &  0.126  &  0.091  & 0.012 & 0.102  & 0.162  \\
\cline{3-3}
&  & 20000  &  &   0.151  &  0.112  &  0.090  & 0.008 & 0.090 & 0.331  \\
\cline{3-3}
&  & 30000  & & 0.138  & 0.063 & 0.068 & 0.007 & 0.069  & 0.492  \\
\cline{2-3}
  & 2 & \multirow{2}{*}{10000} &   & 0.130 &   0.062 &  0.066  &  0.006 &   0.066& 0.196 \\
\cline{2-2}
 & 3 & &   &  0.125  &  0.066 &   0.063  &  0.006&  0.065& 0.223  \\
  \cline{2-3}
 & 10 & 2000 &  &  0.149 &   0.083 &    0.075 &    0.009  &  0.079  & 0.081 \\
\hline
\hline
  \multirow{2}{*}{Adaptive MH } & 10 &  \multirow{2}{*}{2000}&  \multirow{2}{*}{In2}  & 0.158 & 0.082 & 0.087 & 0.012 &  0.084 &  0.090  \\
    \cline{2-2}
  & 100 &  & & 0.146 &  0.076  & 0.073  & 0.010       &  0.076 & 0.634 \\
 \hline
 \hline
  \multirow{3}{*}{Slice}  & 3 & \multirow{2}{*}{2000}  & \multirow{3}{*}{In2}    & 0.204  &   0.105 &   0.103 &   0.022 & 0.108 & 0.156 \\
  % \cline{2-2}
 % & 5 & &   & 0.191 &   0.092 &   0.096 &   0.018 &   0.099  &\\
  \cline{2-2}
  & 10 & & & 0.188  &  0.091  &  0.095   & 0.018 & 0.098 & 0.463 \\
    \cline{2-3}
  & 3 & 10000 &   & 0.132  & 0.051  & 0.066  &  0.007   & 0.064 & 0.783\\
\hline
\end{tabular}
\caption{{\bf (Ex-Sect-\ref{ASM_within_Gibbs_SIMU})} Mean Absolute Error (MAE) in the estimation of four statistics (first component) and normalized computing time. All the techniques are used within a Gibbs sampler: $N_G$ is the number of iterations of the Gibbs sampler whereas $T$ is is the number of iterations of the technique within Gibbs (so that $T \times N_G$ is the global number of MCMC iterations). The best results (in each column, and in each panel) are highlighted with bold-face.}
\label{TableToyExGibbs}
\end{center}
\end{table*}

\end{document}